\documentclass[aps,prl,twocolumn,superscriptaddress]{revtex4-2}
\usepackage{graphicx}
\usepackage{dcolumn}
\usepackage{bm}
\usepackage{amsmath}
\usepackage{amssymb}
\usepackage{latexsym}
\usepackage{epsfig}
\usepackage{amsbsy}
\usepackage{array}
\usepackage{amssymb}
\usepackage{setspace}
\usepackage{float,url}
\usepackage{textcomp}
\usepackage{multirow}
\usepackage{threeparttable}
\usepackage[colorlinks=true,
            linkcolor=blue,
            anchorcolor=blue,
	        urlcolor=blue,
            citecolor=blue]{hyperref}

\begin{document}

\preprint{APS/123-QED}

\title{Universal scaling of higher-order cumulants in quantum  isotropic spin chains
}

\author{Shixian Jiang}
\thanks{These authors contributed equally to this work}
\affiliation{College of Science, National University of Defense Technology, 410073 Changsha, China}%

\author{Jianpeng Liu}
\thanks{These authors contributed equally to this work}
\affiliation{College of Science, National University of Defense Technology, 410073 Changsha, China}%
\affiliation{Hunan Key Laboratory of Extreme Matter and Applications, National University of Defense Technology, Changsha 410073, China}
\affiliation{Hunan Research Center of the Basic Discipline for Physical States, National University of Defense Technology, Changsha 410073, China}

\author{Jianmin Yuan}
\affiliation{Institute of Atomic and Molecular Physics, Jilin University, 130012 Changchun, China}
\affiliation{College of Science, National University of Defense Technology, 410073 Changsha, China}%

\author{Xi-Wen Guan}
\affiliation{Innovation Academy for Precision Measurement Science and Technology, Chinese Academy of Sciences, Wuhan 430071, China}%
\affiliation{NSFC-SPTP Peng Huanwu Center for Fundamental Theory, Xi'an 710127, China}
\affiliation{Department of Fundamental and Theoretical Physics, Research School of Physics, Australian National University, Canberra ACT 0200, Australia}

\author{Yongqiang Li}
\email{li\_yq@nudt.edu.cn}
\affiliation{College of Science, National University of Defense Technology, 410073 Changsha, China}%
\affiliation{Hunan Key Laboratory of Extreme Matter and Applications, National University of Defense Technology, Changsha 410073, China}
\affiliation{Hunan Research Center of the Basic Discipline for Physical States, National University of Defense Technology, Changsha 410073, China}

\date{\today}

\begin{abstract} 
Understanding universal behavior of  far-from-equilibrium transport dynamics  at a quantum many-body level  is a  longstanding challenge.
In particular, a full characterization of  universal dynamics of  nonlocal correlation functions  still remains largely unknown.
In this letter,  we uncover universal  scaling laws of higher-order  cumulants  in one-dimensional  isotropic  Heisenberg model, revealing anomalous behaviors of nonequilibrium dynamics exclusively accessible in higher-order correlations.
By means of  numerical simulations and full counting statistics, we determine  the power laws of both the spin polarization transfer  and contrast cumulants for different kinds of helix and domain-wall initial states. 
Building on  such physical  states,  we unify  the scaling  behavior of the higher-order cumulants, giving rise  to two types of dynamics:  anomalous diffusive and superdiffusive. 
For the former, these higher cumulants show a deviation from Gaussian statistics, with the scaling exponents being identical for the first four orders. 
For the latter, however, we observe a  breakdown of KPZ universality, with the exponents of the third and fourth orders differing significantly from those of the first two. 
Our results are also agreeable with recent experimental observations,  advancing  understanding of far-from-equilibrium transport phenomena.

\end{abstract}

\maketitle

\textit{Introduction}\textemdash 
Characterization of quantum transport  in many-body systems with  charge fluctuations, spin currents, etc., is truly  one of the frontiers in physics~\cite{RevModPhys.94.045006,RevModPhys.81.1665,RevModPhys.83.863,RevModPhys.93.025003,Guan_2022}.
It has been understood in general  that conserved quantities present universal scaling-invariant macroscopic transport, including ballistic, superdiffusive and diffusive propagation, regardless of microscopic scenarios~\cite{wyatt2005quantum,bulchandani2021superdiffusion,PhysRevLett.129.230602,PhysRevLett.121.230602,PhysRevLett.122.127202,PhysRevLett.123.186601,Scheie2021,PhysRevLett.106.220601,10.1073/pnas.1906914116,Gopalakrishnan_2023,PhysRevLett.122.210602}.
Such  emergent many-body hydrodynamics has been widely predicted in quantum systems, even in conditions far from equilibrium, high temperature, or disordered, deepening our understanding of quantum dynamical criticality~\cite{PhysRevLett.56.889,10.1073/pnas.1512261112,Nardis_2022,PhysRevLett.122.150605,PhysRevLett.125.245303,PhysRevX.13.011033,PhysRevLett.132.130401,RevModPhys.86.153,PhysRevLett.134.187101}.  
However, a comprehensive classification of nonequilibrium dynamics necessitates going beyond the  local observations, {\it i.e.} higher-order correlations commonly reveal distinct scaling behavior.      

Higher-order cumulants serve as a vital tool for probing multipoint correlations and revealing fundamental insights into non-Gaussian transport ~\cite{joshi2022observing,PhysRevLett.84.4882}. 
It is generally accepted that ballistic transport universally exhibits linear temporal scaling for all cumulant orders~\cite{10.21468/SciPostPhys.15.4.136,10.21468/SciPostPhys.8.1.007}, whereas diffusive transport manifests order-dependent scaling whose general form remains unresolved~\cite{PhysRevLett.132.017101}.
 As promising platforms for the investigation of  far-from-equilibrium dynamics, the quantum simulations and quantum circuits drastically  leverage single-site addressing to probe multipoint correlations. 
 Recent experimental breakthroughs have been achieved for the study of  elusive higher-order cumulants in  far-from-equilibrium dynamics: e.g. in the chaotic systems, generic initial states exhibit identical diffusive scaling in both first- (density) and second-order cumulants (density fluctuations)~\cite{wienand2024emergence}. 
For the integrable isotropic Heisenberg spin chain, state-dependent transport behavior  is observed in the first-order cumulant, {\it i.e.} the  Kardar-Parisi-Zhang (KPZ) superdiffusion emerges for the domain-wall (DW) initial state~\cite{10.1126/science.abk2397}, in contrast to the diffusive transport for spin-helix (SH) initial state~\cite{jepsen2020spin,PhysRevX.11.041054,jepsen2022long}. 
It  was also evidenced for the DW initial state  that higher-order  cumulants deviate from KPZ superdiffusive dynamics~\cite{rosenberg2024dynamics,valli2024}.
However, this observation of the dynamical universality  for the DW state remains in  debate ~\cite{PhysRevLett.132.017101,PhysRevLett.134.097104,PhysRevLett.131.197102}.
Meanwhile, scaling behavior  of the higher-order cumulants for the SH state is still an open problem. 

\begin{figure*}[ht!]
\centering
\includegraphics[width=7 in,trim=0 0 0 0,clip]{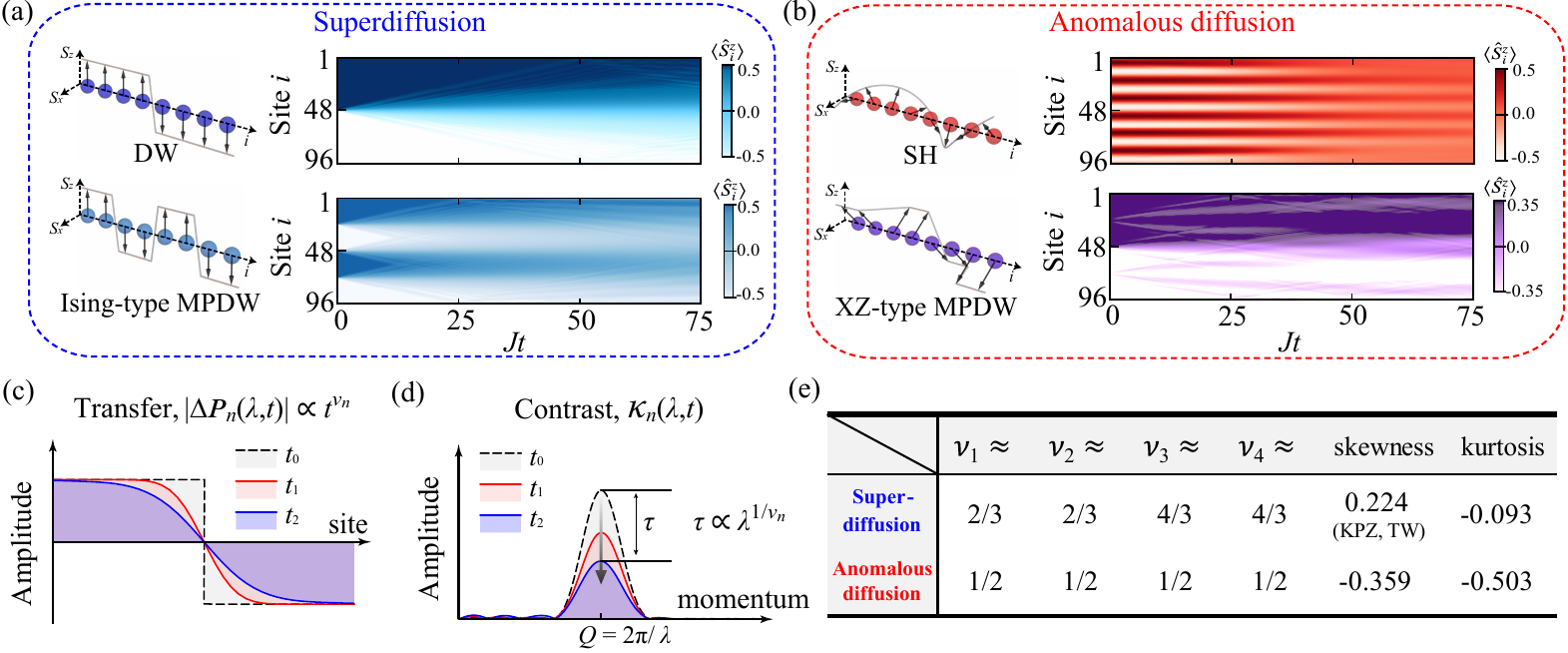}
\caption{Cumulant scaling and its exponents in the isotropic Heisenberg spin chain. Far-from-equilibrium dynamics of spin polarization $\langle \hat{S}^{z}_i \rangle$ for initial states:   (a) DW~\cite{10.1126/science.abk2397,rosenberg2024dynamics} and Ising-type MPDW states,  and (b) SH~\cite{jepsen2020spin,jepsen2022long} and  XZ-type MPDW states, which respectively show  superdiffusive and anomalous diffusive dynamics.
The contour plots illustrate the time evolution of initial real-space spin distributions. (c) shows  $n$th-order cumulant of spin polarization transfer across the domain wall.  (d) illustrates $n$th-order cumulant of the contrast  for a wavevector $Q=2 \pi/\lambda$, where $\nu_n$ denotes the  scaling exponent of $n$th-order cumulant.
(e) shows table of the scaling exponents of first four orders $\nu_{1,2,3,4}$ and standardized cumulants of the third (skewness) and fourth orders (excess kurtosis) for these two types of distinct long-time dynamics obtained by (c) and (d), where TW denotes skewness from the Tracy-Widom distribution in the KPZ class~\cite{tracy1994level}.} 
\label{fig1}
\end{figure*}

In this Letter, we uncover  such miscellaneous scaling behavior of higher-order  cumulants for different experimentally realizable  initial states in the one-dimensional (1D) isotropic Heisenberg chain, typically the  SH~\cite{jepsen2020spin,PhysRevX.11.041054,jepsen2022long} and DW states~\cite{10.1126/science.abk2397,rosenberg2024dynamics}.
Using time-dependent variational principle (TDVP) algorithm and full counting statistics (FCS)~\cite{PhysRevLett.131.210402,PhysRevLett.131.140401,PhysRevLett.133.240403,PhysRevLett.128.160601,PhysRevLett.128.090604,PhysRevLett.132.017101,valli2024}, 
we perform large-scale calculations on the higher-order cumulants,  revealing two distinct types of anomalous quantum transport in experimentally accessible timescales.
For the SH initial state, the non-Gaussian statistics appear in the third- and fourth-order cumulantes and they share the same scaling exponents with the first two.
Whereas for the DW initial state, the third and fourth orders exhibit distinct temporal exponents from the first two, with the standardized cumulant deviating from the KPZ conjecture. 
To solidify this framework, we further study the higher-order cumulants for a multi-periodic domain-wall (MPDW) state with  specific local spin polarization patterns.
Our results confirm the recent experimental observations of the first few cumulants \cite{jepsen2020spin,10.1126/science.abk2397,rosenberg2024dynamics} and open a promising way to identify universal temporal scaling exponents for different initial states in different quantum systems  in current experiments.

\textit{Model and methods}\textemdash 
We study the isotropic 1D Heisenberg spin chain, whose Hamiltonian reads 
\begin{equation}
\hat{H} = -J \sum_i(\hat{S}^x_i\hat{S}^x_{i+1} + \hat{S}^y_i\hat{S}^y_{i+1} + \hat{S}^z_i\hat{S}^z_{i+1}),
\label{eq_hens}
\end{equation}
where $\hat{S}^{x,y,z}_i$ denotes the spin-1/2 operators  at  site $i$, and $J=1$ is the coupling strength.
The Hamiltonian of Eq.~(\ref{eq_hens}) possesses a global SU(2) symmetry, conserving all total spin components and determining macroscopic transport in the long-time scale~\cite{Ljubotina2017}.
Two types of distinct quantum transport behavior are observed in local quantities, {\it i.e.} for the pure DW state, magnetization transfer exhibits the anomalous  scaling of superdiffusion [Fig.~\ref{fig1}(a)(c)]~\cite{10.1126/science.abk2397,rosenberg2024dynamics}, though ultimately recovering diffusion at late times~\cite{PhysRevB.96.195151,ljubotina2017class,misguich2019domain,gamayun2019domain}; for the SH state~\cite{PhysRevLett.113.147205}, diffusive transport manifested in magnetization contrast emerges within experimentally accessible timescales [Fig.~\ref{fig1}(b)(d)]~\cite{jepsen2020spin,jepsen2022long}. Using the  generalized initial MPDW state [Fig.~\ref{fig1}(a)(b)],  we further establish  a unified description of  the nonequilibrium transport dynamics [Fig.~\ref{fig1}(c)(d)]. Detailed definitions of these initial states are provided in Supplemental Material (SM)~\cite{SM}.

To characterize higher-order fluctuations in the far-from-equilibrium transport dynamics, we evaluate $n$th-order cumulants at time $t$ via spin polarization transfer~\cite{Moments_and_cumulants1995,PhysRevC.108.064901}
\begin{equation}
{\mathcal P}_n(\lambda,t) = \sum_{i_1, \cdots, i_n}\langle \hat{\mathcal{S}}_{i_1}^z \hat{\mathcal{S}}_{i_2}^z \cdots \hat{\mathcal{S}}_{i_n}^z \rangle_c, 
\label{eq_cumulant}
\end{equation}
with $\hat{\mathcal{S}}_{i}^z \equiv \text{sgn}[\langle \hat S_i^z(0) \rangle] \hat{S}_{i}^z(t)$, 
and weighted contrast by performing Fourier transform at wavevector $Q=2\pi/\lambda$
\begin{equation}
{\mathcal K}_n(\lambda,t) = \sum_{i_1, \cdots, i_n} \langle \hat{\mathcal{Q}}_{i_1}^z \hat{\mathcal{Q}}_{i_2}^z \cdots \hat{\mathcal{Q}}_{i_n}^z \rangle_c,
\label{eq_contrast},
\end{equation}
where $\lambda$ is wavelength of the initial periodic product state, and 
$\hat{\mathcal{Q}}_{i}^z \equiv -\cos(Q i + \theta) \hat{S}_{i}^z(t)$ with $\theta$ being the spatial phase of the initial state~\cite{SM}. 
$\langle  \cdots  \rangle_c$ denotes the $n$th-order cumulants composed of 
$n$-point operators of $\hat{\mathcal{S}}^z_{i}$ or $\hat{\mathcal{Q}}^z_{i}$~\cite{SM,PhysRevLett.131.197102,PhysRevA.6.1741, PhysRevB.10.265, PhysRevC.107.024910}.
In the long-time limit, cumulants are expected to grow asymptotically as $|\Delta{\mathcal P}_n(\lambda,t)| \equiv |{\mathcal P}_n(\lambda,t)- {\mathcal P}_n(\lambda,0)| \propto t^{\nu_n}$~\cite{Ljubotina2017}, where $\nu_n$ is the scaling exponent for the $n$th order [Fig.~\ref{fig1}(c)(e)]. 
The threshold time $\tau$, extracted from ${\mathcal K}_n(\lambda,t)$ either when it decays to 0.4 ($n=1$ for local quantities) or reaches its peak position ($n=2$, 3 or 4 for nonlocal quantities), obeying  the scaling law of  $\tau \propto \lambda^{1/\nu_n}$ for different $\lambda$ {[Fig.~\ref{fig1}(d)(e)]~\cite{jepsen2020spin}. 
To transform dissimilar features into the same scale~\cite{rosenberg2024dynamics}, we introduce standardized cumulants, including skewness $\gamma_3(\lambda,t)$ and excess kurtosis $\gamma_4(\lambda,t)$~\cite{10.1126/science.abk2397}. 
Note here that the first four cumulants represent the mean, variance, asymmetry, and peakedness of statistical distributions, respectively~\cite{Moments_and_cumulants1995}.

In order to obtain scaling exponents of the cumulants in  the long-time dynamics, we employ TDVP method in terms of  the matrix product state framework and  the ITensor library~\cite{10.21468/SciPostPhysCodeb.4,PhysRevLett.107.070601,PhysRevB.99.054307}.
Specifically, for the balance of finite-size effects and computational cost, the hybrid two-site and one-site TDVP method is applied~\cite{PhysRevB.99.054307,PhysRevLett.130.246402,PhysRevB.102.094315}. 
To further validate our approach, we also present higher-order cumulants using both FCS and subsystem spin fluctuations. Numerical details and convergence tests are presented in SM~\cite{SM}.

\begin{figure}[t!]
\centering
\includegraphics[width=3.5 in,trim=0 0 0 0,clip]{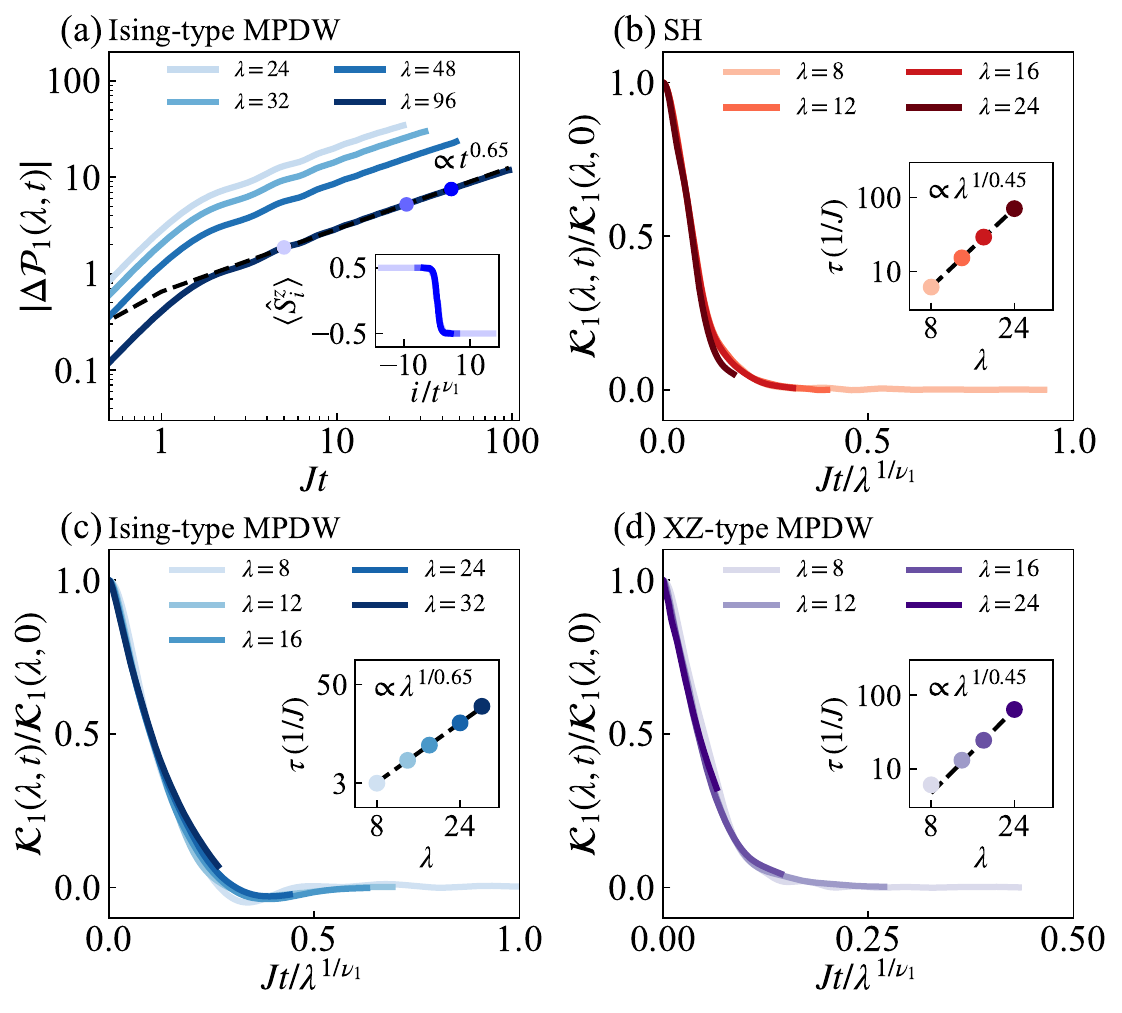}
\caption{First-order cumulant of the isotropic Heisenberg chain.
(a) shows spin polarization transfer $| \Delta{\mathcal P}_1(\lambda, t) |$ for Ising-type (superdiffusive) initial states with wavelength $\lambda=24$ to 96. 
The inset shows rescaled spin distributions for $\lambda=96$ at time $Jt=5$, $25$, and $45$. 
(b-d) Rescaled contrast curve ${\mathcal K}_1(\lambda,t)$ for (b) SH (diffusive), (c) Ising-type MPDW (superdiffusive), and (d) XZ-type MPDW (diffusive) initial states. 
The insets show the relaxation time  $\tau$ following a power-law scaling $\tau \propto \lambda^{1/\nu_1}$.
For Ising-type initial states, both spin polarization transfer and contrast cumulants  yield consistent superdiffusive dynamics  with the scaling exponent $\nu_1 \approx 0.65$, while for the XZ-type initial states, diffusive dynamics emerges with $\nu_1 \approx 0.45$. Lattice size is $L=96$.}
\label{fig2}
\end{figure}

\textit{First-order cumulant via unified description}\textemdash 
We first  analyze  the first-order cumulant in far-from-equilibrium transport dynamics of the quantum isotropic Heisenberg chain. 
Three initial states, including DW, MPDW, and SH states, are investigated to extract scaling behavior in the spin transport. 
The many-body dynamics is quantified through spin polarization transfer $|\Delta{\mathcal P}_1(\lambda,t) |$ and contrast decay ${\mathcal K}_1(\lambda,t)$ in  the experimentally accessible timescales with $Jt\approx 100$.

As expected, for  the DW initial state, the transport dynamics exhibits superdiffusive behavior. 
Fig.~\ref{fig2}(a) shows first-order spin polarization transfer for the DW initial  state (with $\lambda=L=96$).
The long-time dynamics of the spin polarization transfer exhibits  a temporal power-law growth, $i.e.$ $|\Delta{\mathcal P}_1(\lambda,t)| \propto t^{\nu_1}$ with the scaling exponent $\nu_1 \approx 0.65$, confirming prior numerical results ~\cite{PhysRevB.96.195151,ljubotina2017class,misguich2019domain,gamayun2019domain} and experimental observations~\cite{10.1126/science.abk2397} for the DW initial state.
The data collapse of the  real-space spin distributions shows a consistency of the  rescaling with $\nu_1 =0.65$ [inset of Fig.~\ref{fig2}(a)]. 
In contrast, for the  SH initial state, we  observe  that a diffusive transport emerges in the long-time dynamics, see the scaling collapse of contrast decay ${\mathcal K}_1(\lambda,t)$ in Fig.~\ref{fig2}(b) and the  power-law scaling of relaxation time $\tau \propto \lambda^{1/\nu_1}$ in the inset of {Fig.~\ref{fig2}(b)}, and also see discussions in ~\cite{PhysRevLett.132.220404,PhysRevB.104.L081410,PhysRevA.106.043306,PhysRevA.110.023313,PhysRevB.107.235408,PhysRevB.108.075135,PhysRevB.103.115435,PhysRevB.104.195409,PhysRevB.107.214422}. 
We note that the resulting exponent $\nu_1 \approx 0.45$ for the SH initial state agrees with the experimental observations~\cite{jepsen2020spin}. 
Nevertheless the  boundary effects  do  disrupt the scaling behavior in a finite system.
Here we  find that a lattice size of $L=96$ is sufficient enough for our study, see  SM~\cite{SM}. 

In order to  provide a unified description for experimentally relevant transport dynamics, we further consider  the MPDW  initial states to reproduce transport scaling behavior in the isotropic Heisenberg chain. 
For a Ising-type MPDW initial state, {\it i.e.} multi-periodic DW state [Fig.~\ref{fig1}(a)],  we obtain the superdiffusive scaling exponent $\nu_1 \approx 0.65$ of the first-order cumulant through spin polarization transfer  [Fig.~\ref{fig2}(a)] and contrast decay [Fig.~\ref{fig2}(c)]. 
Meanwhile we reconfigure the initial state by adjusting the local spin mixture between $S^x_i$ and $S^z_i$~\cite{SM} to reproduce the diffusive transport observed in the SH state. 
Starting from this modified initial state, {\it i.e.} XZ-type MPDW state [Fig.~\ref{fig1}(b)], we indeed further observe a diffusive transport emerging  with $\nu_1\approx 0.45$ in the long-time dynamics, see the  scaling collapse of contrast decay [Fig.~\ref{fig2}(d)] and power-law scaling of relaxation time [inset of Fig.~\ref{fig2}(d)]. 
The consistency of the scaling exponents of the spin polarization transfer and the contrast for different initial states establishes a unified framework for understanding  the far-from-equilibrium transport dynamics. 

\begin{figure}[t!]
\centering
\includegraphics[width=3.5 in]{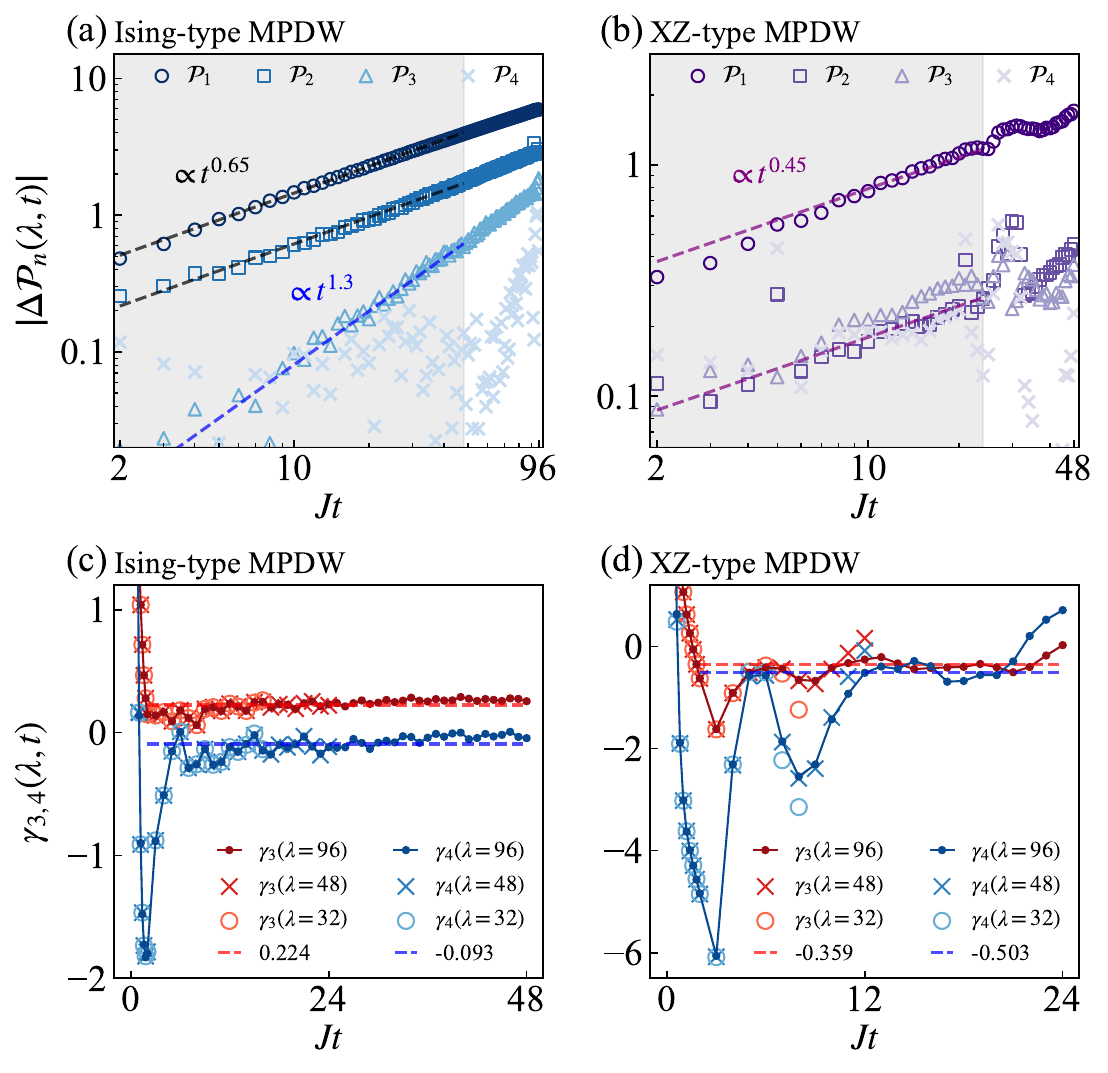}
\caption{Higher-order cumulants  of spin polarization transfer $| \Delta{\mathcal P}_n(\lambda, t)|$ for Ising-type (superdiffusive) and XZ-type (diffusive) initial states.  (a)(c) For Ising-type MPDW states, (a) third- and fourth-order cumulant exponents  ($\nu_{3,4} \approx 1.3$) are twice those of the first and second orders $\nu_{1,2} \approx 0.65$ with $\lambda=96$ as an example, and (c) standardized cumulants confirm deviation from KPZ universality through skewness $\gamma_3\approx 0.224$~\cite{PhysRevLett.84.4882} and excess kurtosis $\gamma_4 \approx -0.093$~\cite{rosenberg2024dynamics}.
(b)(d) For the XZ-type MPDW states, (b) identical scaling exponent with $\nu_n \approx 0.45$ ($n=1, 2, 3, 4$), and (d) non-Gaussian statistics with $\gamma_3 \approx -0.359$ and $\gamma_4 \approx -0.503$. Lattice size is $L=96$.
 } 
\label{fig3}
\end{figure}

\textit{Higher-order cumulants via spin polarization transfer}\textemdash 
We now turn to main conclusions about higher-order cumulants emerged in far-from-equilibrium transport dynamics. 
The  classical Heisenberg spin chain  initiated from an infinite-temperature equilibrium state displays  the scaling relation $\nu_{2n} = n\nu_2$ ($n=1,\,2,\,3$) for both superdiffusive ($\nu_2=2/3$) and diffusive ($\nu_2=1/2$) regimes~\cite{PhysRevLett.132.017101}.
For the finite-temperature initial states, our analysis of the quantum Heisenberg chain reveals two types of fluctuation dynamics: superdiffusive  dynamics with $\nu_{3,4}=2\nu_{1,2}$, and diffusive one with  $\nu_{1,2}=\nu_{3,4}$.
The later is a reminiscence of the chaotic dynamics ~\cite{PhysRevLett.131.210402}.

Fig.~\ref{fig3} illustrates higher-order polarization transfer $|\Delta{\mathcal P}_n(\lambda,t)|$ for the  Ising-type MPDW states (a) and for the  XZ-type MPDW states (b) in the isotropic Heisenberg spin chain.
Here, the gray shaded regions  correspond to the analysis domains: $Jt < \lambda/2$ for Ising-type state, and $Jt < \lambda/4$ for XZ-type MPDW state~\cite{SM}. %
We observe that the Ising-type initial state (superdiffusive) yields $|\Delta{\mathcal P}_{1,2} (\lambda, t)| \propto t^{0.65}$, and $|\Delta{\mathcal P}_{3,4} (\lambda, t)| \propto t^{1.3}$ [Fig.~\ref{fig3}(a)], reminiscent of previous theoretical predictions for the infinite-temperature Heisenberg model~\cite{PhysRevLett.132.017101}.
In contrast to that, the XZ-type initial state (diffusive) demonstrates distinct behavior of the polarization transfer  $|\Delta{\mathcal P}_{1,2,3,4} (\lambda, t)| \propto t^{0.45}$, see [Fig.~\ref{fig3}(b)].
Our results show that there exist two types of transport belonging to different dynamical universality classes: superdiffusive and diffusive. 
Here we remark that long-time jitter and divergence in fourth-order cumulant stem from boundary-induced disturbances~\cite{rosenberg2024dynamics}. 
Extended finite-size analysis is provided in SM~\cite{SM}.

Moreover, we find that deviations from KPZ superdiffusion and normal diffusion are exclusively resolved by the standardized higher-order cumulants.
Fig.~\ref{fig3}(c)(d) shows the skewness $\gamma_3(\lambda,t)= \Delta{\tilde{\mathcal P}}_3(\lambda,t)/\Delta{\tilde{\mathcal P}}_2^{3/2}(\lambda,t)$ and excess kurtosis $\gamma_4(\lambda,t)=\Delta{\tilde{\mathcal P}}_4(\lambda,t)/ \Delta{\tilde{\mathcal P}}_2^{2}(\lambda,t)$ in the transport dynamics [shaded regions in Fig.~\ref{fig3}(a)(b)].
Here, the averaged transfer $\Delta{\tilde{\mathcal P}}_n(\lambda,t) \equiv \Delta{\mathcal P}_n(\lambda,t) / n_c$ with $n_c\equiv 2L/\lambda-1$ counting the up-down edge points of initial DW structure, e.g. $n_c=3$ for double-periodic DW state. 
We find that the Ising-type MPDW states [Fig.~\ref{fig3}(c)] result in $\gamma_3\approx 0.224$ and $\gamma_4 \approx -0.093$ at finite timescales, where these values agree with experimental observations~\cite{rosenberg2024dynamics} and the latter deviates from the KPZ dynamics~\cite{baik2000limiting,tracy1994level}.
Moreover, the XZ-type MPDW states [Fig.~\ref{fig3}(d)] give rise to the distinctive non-Gaussian statistics with $\gamma_{3} \approx -0.359$ and $\gamma_4\approx-0.503$, providing anomalous diffusive transport in the integrable quantum system~\cite{PhysRevB.109.024417}.
Note here that the late-stage increase in skewness and excess kurtosis is also a result of finite-size effects, see SM~\cite{SM}. 

\begin{figure}[t!]
\centering
\includegraphics[width=3.4 in]{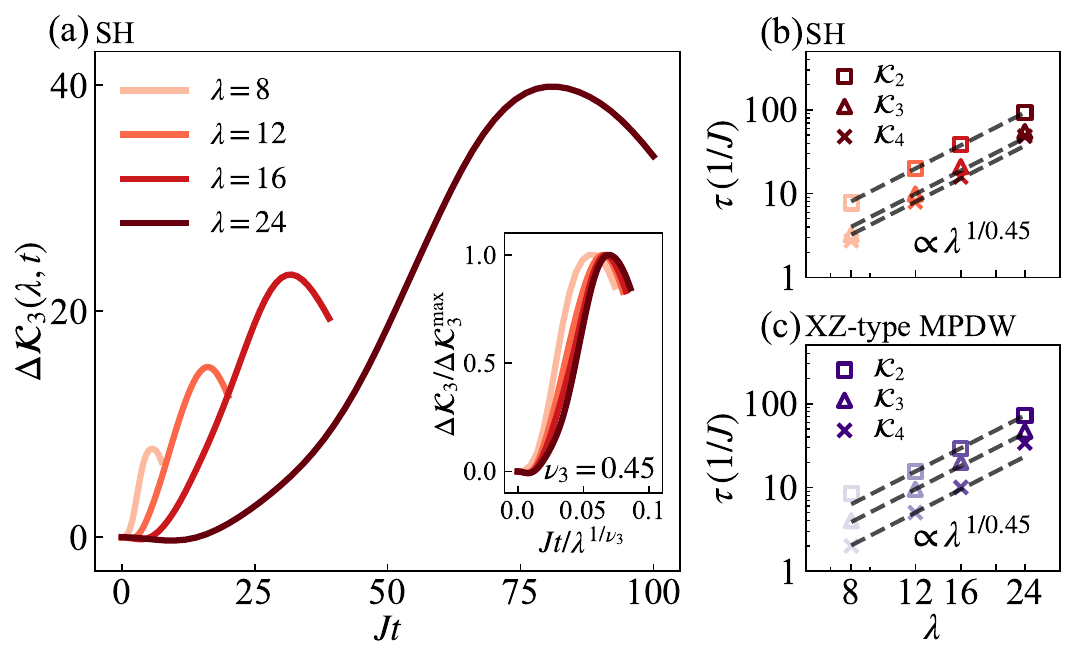}
\caption{Higher-order cumulants via contrast $\mathcal{K}_n(\lambda, t)$ for XZ-type MPDW (diffusive) and SH (diffusive) initial states. (a) shows contrast dynamics of the third-order cumulant ${\mathcal K}_3(\lambda,t)$ for the SH states with different wavelengths. The inset shows data collapse of rescaled contrast from (a), where $\Delta{\mathcal K}_3 \equiv {\mathcal K}_3(\lambda,t) - {\mathcal K}_3(\lambda,0)$ and $\Delta{\mathcal K}^{\rm max}_3$ denotes the first maximum. 
(b)(c) Power-law scaling of the growth threshold time $\tau$ when the cumulant reaches its first maximum, versus wavelength $\lambda$ for the second- ($\square$), third- ($\triangle$) and fourth-order ($\times$) cumulants for different initial states: (b)  SH and (c) XZ-type MPDW states, respectively. 
Scaling exponents from contrast agree with those obtained via spin polarization transfer. Lattice size is $L=96$.}
\label{fig4}
\end{figure}

In order to  further verify the above conclusions, we employ FCS to characterize spin transport fluctuations. 
This method extracts the cumulants from the quantum generating function of the  spin polarization transfer~\cite{RevModPhys.81.1665,10.1126/science.abk2397,valli2024}.
The detailed formulas and dynamical scalings for Ising- and XZ-type MPDW states are given in SM~\cite{SM}.
In addition, quantum fluctuations versus subsystem sizes~\cite{wienand2024emergence} provides an alternative tool for extracting scaling exponents of the higher-order cumulants, and more discussions are given in SM~\cite{SM}.

\textit{Higher-order cumulants from contrast}\textemdash 
To further examine transport behavior in higher-order cumulants, we extract scaling exponents from the contrast ${\mathcal K}_n(\lambda,t)$ in the far-from-equilibrium dynamics of the Heisenberg chain. 
For both XZ-type MPDW and SH initial states, time evolution generates spatial correlations between sites, producing finite higher-order cumulants. Fig.~\ref{fig4}(a) shows the third-order contrast ${\mathcal K}_3$ for the SH initial state, where we observe that it rises to a maximum at time $\tau$ before decaying. As expected, the third-order cumulant exhibits scaling collapse when rescaled by $\lambda^{1/\nu_3}$,  see inset of Fig.~\ref{fig4}(a). 
The scaling exponent $\nu_3$ is identical to the prediction from the first-order contrast decay with $\nu_1 = \nu_3$ [Fig.~\ref{fig2}(b)]. 
A discussion on the rescaled dynamics of ${\mathcal K}_{2,4}$ is provided in SM~\cite{SM}. 

In addition, the growth threshold time $\tau$ for the contrast $\mathcal{K}_{2,3,4}$ reaching their respective first maxima also exhibits power-law scaling $\tau \propto \lambda^{1/\nu_n}$. In Fig.~\ref{fig4}(b)(c), we plot the threshold time $\tau$ versus wavelength $\lambda$ for the SH (b) and the XZ-type MPDW states (c), confirming that the scaling of cumulant dynamics exhibit identical exponents $\nu_{1,2,3,4} \approx 0.45$  for both SH and XZ-type MPDW states. 
The consistency with the scaling exponents obtained from spin polarization transfer measurement $|\Delta{\mathcal P}_n|$ [Fig.~\ref{fig3}(b)] and contrast analysis ${\mathcal K}_n$ [Fig.~\ref{fig4}(b)(c)] validates the reliability of characteristic of the  higher-order cumulants in the far-from-equilibrium dynamics.

\textit{Summary and outlook}\textemdash 
We have established  a theoretical framework for capturing higher-order universal scaling of  far-from-equilibrium transport dynamics in the 1D quantum Heisenberg model. 
Using numerical simulations and full counting statistics, we have obtained the scaling  laws of higher-order  cumulants for different kinds of SH and DW initial states that have not been characterized so far. 
Such  scaling  behavior of the higher-order cumulants, {\it i.e.} the spin polarization transfer  and contrast cumulants for various experimentally accessible initial states, has been classified into  two types of dynamics: anomalous diffusive and superdiffusive. 
The former shows the same exponents of the first four cumulants, whereas the latter indicates that exponents of the third and fourth orders are significantly  different from that of the first two.  
Our results are agreeable with recent experiments~\cite{jepsen2020spin,10.1126/science.abk2397,rosenberg2024dynamics}, and further reveal the essential roles of higher-order nonlocal correlations in understanding far-from-equilibrium transport dynamics in a wide range of systems, such as ultracold  atoms and superconducting compounds ~\cite{10.1126/science.abk2397,jepsen2020spin,rosenberg2024dynamics} etc. 

The obtained anomalous dynamics resolves the open 
question whether the higher-order cumulants exhibit universal scaling laws  in the two classes of diffusive and superdiffusive 
 expansions  toward asymptotic diffusion~\cite{PhysRevB.96.195151,ljubotina2017class,misguich2019domain,gamayun2019domain}. 
Furthermore, our method can be extended to the study of the long-time dynamics in low-dimensional doped strongly correlated fermionic systems, exploring how higher-order cumulants capture the essence of the spin-charge separation phenomenon~\cite{Giamarchi2003,Senaratne2022}.

\textit{Acknowledgments}\textemdash 
We acknowledge useful discussions
with Zhensheng Yuan, Lei Feng, and Xuchen Yang. Y.L. acknowledges the support by the National Natural
Science Foundation of China (Grants No. 12074431, No. 12374252 and No. 12274384), and
the Science and Technology Innovation Program of Hunan
Province under Grant No. 2024RC1046. 
J.L. acknowledges the support by the National Natural
Science Foundation of China (Grants No. 12404332). X.W.G. is supported by the NSFC key grant No. 12134015, the NSFC  key grant  No. 92365202  and the Innovation Program for Quantum Science and Technology 2021ZD0302000.

\bibliography{ref}

\clearpage

\onecolumngrid

\setcounter{figure}{0}
\setcounter{table}{0}
\setcounter{equation}{0}
\renewcommand{\thefigure}{S\arabic{figure}}
\renewcommand{\thetable}{S\arabic{table}}
\renewcommand{\theequation}{S\arabic{equation}}

\centerline{\Large{\bf Supplemental Material: }}
\centerline{\Large{\bf Universal scaling of higher-order cumulants in quantum  isotropic spin chains}}

\maketitle

\renewcommand{\theenumi}{\Roman{enumi}}  %
\renewcommand{\theenumii}{\Alph{enumii}} %
\renewcommand{\theenumiii}{\arabic{enumiii}} %

\section*{Contents}
\addcontentsline{toc}{chapter}{Table of Contents}
\begin{enumerate}
    \item \hyperref[chapter:Doh]{Higher-order cumulants within full counting statistics\hfill \pageref{chapter:Doh}}
    \begin{enumerate}
        \item \hyperref[section:Qgf]{Quantum generating function\hfill \pageref{section:Qgf}}
        \item \hyperref[section:Eoh]{Expansions of higher-order moments and cumulants\hfill \pageref{section:Eoh}}
    \end{enumerate}

    \item \hyperref[chapter:Pot]{Properties of the initial states \hfill \pageref{chapter:Pot}}
    \begin{enumerate}
        \item \hyperref[section:Dot]{Definitions of the initial states \hfill \pageref{section:Dot}}
        \item \hyperref[section:Pso]{Parameter sensitivity of multi-periodic states \hfill \pageref{section:Pso}}
    \end{enumerate}

    \item \hyperref[chapter:Ctd]{Convergence tests in the far-from-equilibrium dynamics \hfill \pageref{chapter:Ctd}}
    \begin{enumerate}
         \item \hyperref[section:Cos]{ Cumulant dynamics of spin polarization transfer\hfill \pageref{section:Cos}}
        \item \hyperref[section:Coc]{ Cumulant dynamics of contrast analysis \hfill \pageref{section:Coc}}
    \end{enumerate}

    \item \hyperref[chapter:Ado]{Additional data of spin polarization transfer and contrast cumulants \hfill \pageref{chapter:Ado}}
    \begin{enumerate}
        \item \hyperref[section:Spt]{Spin polarization transfer scaling for Ising-type and XZ-type MPDW states\hfill \pageref{section:Spt}}
        \item \hyperref[section:Csi]{Contrast scaling for SH and XZ-type MPDW states \hfill \pageref{section:Csi}}
    \end{enumerate}

    \item \hyperref[chapter:Smf]{Dynamical scaling of subsystem fluctuations\hfill \pageref{chapter:Smf}}

\end{enumerate}

\vspace{1em}

This supplemental material provides comprehensive supporting details for the universal scaling of higher-order cumulants in quantum isotropic spin chains. 
Section I derives the higher-order cumulants in the framework of full counting statistics (FCS). 
Section II defines the initial states and examines the dependence of the dynamics on initial phase and wavelength periods.
Section III validates the convergence of the first few cumulants of spin polarization transfer and contrast against system parameters.
Section IV provides extended datasets for Figs. 3 and 4 of the main text. 
Finally, Section V introduces a scaling method for higher-order cumulants based on subsystem fluctuations.

\section{I. Higher-order cumulants within full counting statistics}\label{chapter:Doh}
This section systematically presents the derivation of higher-order moments and cumulants in the framework of FCS, which is widely used in the statistical distributions of transport. 
Within this method, one generally utilizes the quantum generating function (QGF) approach, where its $n$th-order derivatives directly yield the $n$th-order cumulants for various initial states. 
Here, we find the method yields results in agreement with those from time-dependent variational principle (TDVP) algorithm. To analyze the dynamics, we define $\hat{O} \equiv \hat{O}(t)$, which represents a time-dependent operator with conserved global quantities (e.g., charge, spin, or particle density).

\subsection{Quantum generating function }\label{section:Qgf}
In quantum systems, the QGF for cumulants characterizes the statistical distribution of a local observable $\hat{O}$~\cite{PhysRevB.92.115103}
\begin{equation}\label{qgf_G}
G(a) =  \ln \langle e^{a \hat{O}} \rangle,
\end{equation}
where $a = |a|e^{i\phi_a}$ is a complex-valued counting field with amplitude $|a|$ and phase $\phi_a$. 
In this work, we focus on the $z$-component spin distribution $\hat{O} = \sum_{i \in \mathcal{A}} \hat{S}_i^z$, with $\mathcal{A}$ denoting lattice sites.

Cumulants $\kappa_n$ are obtained by calculating the $n$th-order derivative of QGF at $a=0$
\begin{equation}
\kappa_n = \left.\frac{\partial^n}{\partial a^n} G(a)\right|_{a=0},
\label{eq:QGF}
\end{equation}
where the Taylor expansion of $G(a)$ about $a=0$ is given by~\cite{PhysRevC.108.064901}
\begin{equation}
G(a) = \sum_{n=0}^{\infty} \frac{a^n}{n!} \kappa_n = 1 + a\kappa_1 + \frac{a^2}{2!}\kappa_2 + \cdots.
\end{equation}
The complete set of $\kappa_n$, known as the FCS, provides a comprehensive description of quantum fluctuations beyond the first-order expectation values $\langle \hat{O} \rangle$. Detailed forms and physical interpretations of $\kappa_n$ are presented in the following subsection. 

\begin{figure}[t!]
    \centering
    \includegraphics[width=0.6\textwidth]{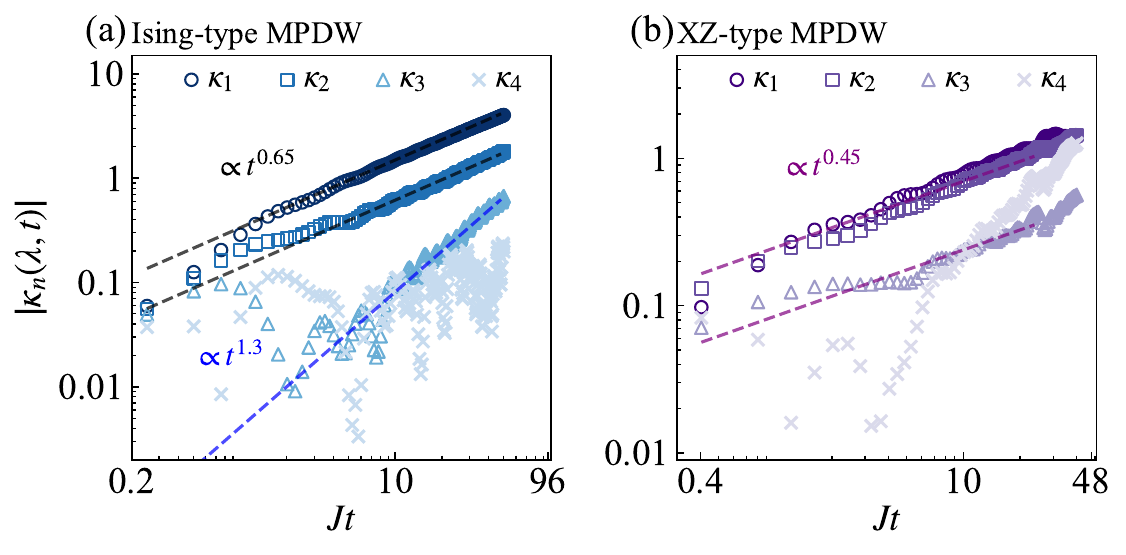}
    \caption{
    Higher-order cumulants for (a) Ising-type and (b) XZ-type MPDW initial states with wavelength $\lambda=96$, obtained by the QGF approach. %
    Dashed lines indicate power-law guides $|\kappa_n(\lambda, t)| \propto t^{\nu_n}$ with scaling exponents $\nu_{1,2} \approx 0.65$ and $\nu_{3,4} \approx 1.3$ for the Ising-type MPDW state, and $\nu_{1, 2, 3, 4}\approx0.45$ for the XZ-type MPDW state. We choose the system size $L=96$.
}
    \label{fig:Sfig_QGF}
\end{figure}
To validate our TDVP findings in the main text, Fig.~\ref{fig:Sfig_QGF} illustrates cumulants $|\kappa_n(\lambda, t)|$ of the isotropic Heisenberg spin chain for the Ising-type and XZ-type multi-periodic domain-wall (MPDW) states, obtained from QGF [Eq.~(\ref{eq:QGF})]. Note here that simulations are performed by gate operations on matrix product state (MPS)~\cite{valli2024} of Eq.~(\ref{qgf_G}) in ITensor library.
Despite temporal fluctuations in the far-from-equilibrium transport dynamics, this different approach confirms the state-dependent emergence of scaling laws in the first four cumulants:
Higher-order scaling exponents exhibit twice the values of the first two orders for the Ising-type MPDW state, whereas scaling exponents remain identical across all orders for the XZ-type MPDW state.
These results are consistent with our TDVP results in Fig.~3 of the main text.

\subsection{Expansions of higher-order moments and cumulants}\label{section:Eoh}
This part derives explicit expressions for higher-order moments and cumulants. Central moments measure deviations of observables from their mean values. The $n$th-order central moment is defined as~\cite{Moments_and_cumulants1995}
\begin{equation}
\begin{aligned}
\mu_n &= \langle ( \hat{O} -\langle \hat{O} \rangle)^n \rangle \\
&= \sum_{k=0}^n \binom{n}{k} (-1)^{n-k} \langle \hat{O}^k \rangle \langle \hat{O} \rangle^{n-k},
\label{eq:n_central_moments}
\end{aligned}
\end{equation}
with explicit expansions of the first four orders given by
\begin{equation}
\begin{aligned}
\mu_1 &= 0, \\
\mu_2 &= \langle \hat{O}^2 \rangle - \langle \hat{O} \rangle^2, \\
\mu_3 &= \langle \hat{O}^3 \rangle - 3\langle \hat{O}^2 \rangle \langle \hat{O} \rangle + 2 \langle \hat{O} \rangle^3, \\
\mu_4 &= \langle \hat{O}^4 \rangle - 4\langle \hat{O}^3 \rangle \langle \hat{O} \rangle + 6\langle \hat{O}^2 \rangle \langle \hat{O} \rangle^2 - 3\langle \hat{O} \rangle^4.
\end{aligned}
\end{equation}
The first four central moments provide a comprehensive characterization of symmetry and tail pattern:
\begin{itemize}
    \item The first-order moment $\mu_1$ gives the mean of distribution.
    \item The second-order moment $\mu_2$ quantifies variance (fluctuations).
    \item The third-order moment $\mu_3$ determines skewness (asymmetry).
    \item The fourth-order moment $\mu_4$ characterizes kurtosis (peakedness).
\end{itemize}

On the other side, the first four cumulants take the following explicit forms obtained from the QGF approach~\cite{Moments_and_cumulants1995,PhysRevC.108.064901}
\begin{equation}
\begin{aligned}
\kappa_1 &= \langle \hat{O} \rangle, \\
\kappa_2 &= \langle \hat{O}^2 \rangle - \langle \hat{O} \rangle^2, \\
\kappa_3 &= \langle \hat{O}^3 \rangle - 3\langle \hat{O}^2 \rangle \langle \hat{O} \rangle + 2 \langle \hat{O} \rangle^3, \\
\kappa_4 &= \langle \hat{O}^4 \rangle - 4\langle \hat{O}^3 \rangle \langle \hat{O} \rangle - 3\langle \hat{O}^2 \rangle^2 + 12\langle \hat{O}^2 \rangle \langle \hat{O} \rangle^2 - 6 \langle \hat{O} \rangle^4.
\end{aligned}
\label{eq:n_cumulants}
\end{equation}
One can find that the relations between central moments $\mu_n$ and cumulants $\kappa_n$, where the first four orders are given by~\cite{Moments_and_cumulants1995}
\begin{equation}
\begin{aligned}
\mu_1 &= 0, \\
\mu_2 &= \kappa_2, \\
\mu_3 &= \kappa_3 , \\
\mu_4  &=  \kappa_4 + 3\mu_2^2.
\end{aligned}
\end{equation}
Note that the first-order central moment is always zero, while the fourth-order central moment equals the fourth-order cumulant plus the square of the second-order central moment.

In addition, standardized cumulants capture key distributional features by scaling higher-order ones with powers of the variance. 
This dimensionless formula provides a clear representation of the non-Gaussian distribution in high-order fluctuations.
Specifically, skewness, defined as $\gamma_3= \kappa_3 / \kappa_2^{3/2}$, quantifies distributional asymmetry ($\gamma_3=0$: symmetry). Excess kurtosis, defined as $\gamma_4= \kappa_4 / \kappa_2^{2}$, measures tail heaviness relative to Gaussian distribution ($\gamma_4=0$: normal kurtosis).
Note here that excess kurtosis can also be expressed equivalently in terms of central moments as  $\mu_4/\mu_2^2 = \gamma_4 + 3$.

We now define the cumulants of the spin polarization transfer and contrast used in the main text. 
The $n$th-order cumulant of spin polarization transfer is defined as
\begin{equation}
    \mathcal{P}_n(\lambda,t) = \sum_{i_1, \cdots, i_n} \langle \hat{\mathcal{S}}_{i_1}^z \hat{\mathcal{S}}_{i_2}^z \cdots \hat{\mathcal{S}}_{i_n}^z \rangle_c,
\label{eq:transfer}
\end{equation}
where the explicit expression of $\langle  ... \rangle_c$  is given by Eq.~(\ref{eq:n_cumulants}). 
For example, the first two orders read
\begin{equation}
\begin{aligned}
    \mathcal{P}_1(\lambda,t) &= \sum_{i_1} \langle \hat{\mathcal{S}^z_{i_1}} \rangle, \\
    \mathcal{P}_2(\lambda,t) &= \sum_{i_1, i_2} \langle \hat{\mathcal{S}^z_{i_1}}\hat{\mathcal{S}^z_{i_2}} \rangle - \sum_{i_1,i_2}  \langle \hat{\mathcal{S}^z_{i_1}} \rangle \langle \hat{\mathcal{S}^z_{i_2}} \rangle.
\end{aligned}
\end{equation}
Here, the local operator $\hat{\mathcal{S}}_{i}^z$ quantifies the spin polarization while incorporating the sign of the initial magnetization
\begin{equation}
\hat{\mathcal{S}}_{i}^z \equiv \mathrm{sgn}[\langle \hat{S}_i^z(0) \rangle] \hat{S}_{i}^z(t), 
\label{eq:sgn_Sz}
\end{equation}
where the sign function is given by
\begin{equation}
\mathrm{sgn}[\langle \hat{S}_i^z(0) \rangle]  =
\begin{cases}
+1, & \langle \hat{S}_i^z(0) \rangle > 0 \\
-1, & \langle \hat{S}_i^z(0) \rangle < 0.
\end{cases}
\end{equation}
Here, the local operator $\langle \hat{\mathcal{S}}_{i}^z\rangle$ maintains uniform orientation for all the sites. 
For example, to avoid the cancellation in odd-order terms for the single-periodic domain-wall (DW) initial state, the first-order cumulant  across all lattice sites is given by $\mathcal{P}_1(\lambda,t) =  \sum_{i\in\mathcal{L}} (+1)\langle \hat S_i^z(t)\rangle + \sum_{i\in\mathcal{R}} (-1)\langle \hat S_i^z(t)\rangle \equiv \sum_i \langle \hat{\mathcal{S}^z_i} \rangle$. 
Here, $\mathcal{L}$ denotes the left half of the chain with spin up, and $\mathcal{R}$ the right half with spin down.
To extract the scaling exponent in the long-time evolution, we focus on the cumulant of transferred spin polarization relative to the initial state, defined as $|\Delta{\mathcal P}_n(\lambda,t)| \equiv |{\mathcal P}_n(\lambda,t)- {\mathcal P}_n(\lambda,0)|$.

The contrast cumulant at wavevector $Q=2\pi/\lambda$ for the multi-periodic state is defined as
\begin{equation}
\mathcal{K}_n(\lambda,t) = \sum_{i_1, \cdots, i_n} \langle \hat{\mathcal{Q}}_{i_1}^z \hat{\mathcal{Q}}_{i_2}^z \cdots \hat{\mathcal{Q}}_{i_n}^z \rangle_c, 
\label{eq:contrast}
\end{equation}
where $\lambda$ is the wavelength of the initial state, and the weighted spin operator $\hat{\mathcal{Q}}_{i}^z \equiv -\cos(Q i + \theta) \hat{S}_{i}^z(t) $ denotes a Fourier transform from real space to momentum space, with $\theta$ denoting the initial phase of cosine-modulated $\langle \hat{S}_i^z(0) \rangle$. 
The explicit expansion of $\langle  \hat{\mathcal{Q}}_{i_1} \hat{\mathcal{Q}}_{i_2} \cdots \hat{\mathcal{Q}}_{i_n}  \rangle_c$
follows the cumulant formulation given in Eq.~(\ref{eq:n_cumulants}), analogous to that for spin polarization transfer. 
Unless specified otherwise, the site $i_n$ in the sum $\sum_{i_1,...,i_n}$ ranges over all lattice sites from $1$ to $L$ for operators $\hat{\mathcal{S}}_{i_n}^z$ and $\hat{\mathcal{Q}}_{i_n}^z$.

\begin{figure}[t!]
    \centering
    \includegraphics[width=1 \textwidth]{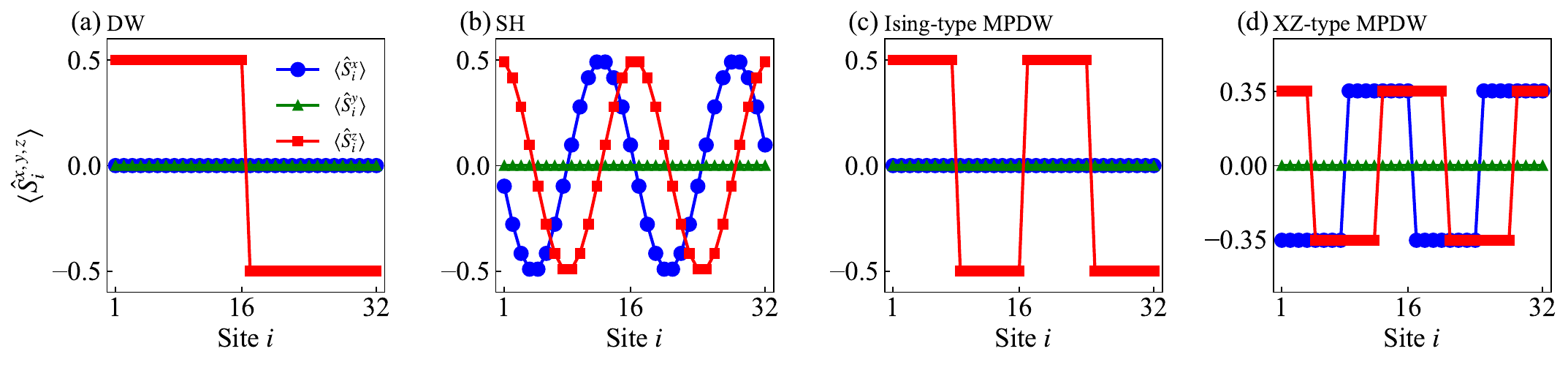}
    \caption{Real-space spin distributions for the different initial states: (a) DW state with $\lambda=32$, (b) SH state with wavelength $\lambda=16$ and initial phase $\theta=0$, (c) Ising-type MPDW state with $\lambda=16$, amplitude ratio $\eta=0$ and $\theta=-\pi/2$, and (d) XZ-type MPDW state with $\lambda=16$, $\eta=1$, $\theta=0$ and phase difference $\phi=\pi/2$. 
    The Ising-type MPDW state recovers the superdiffusive behavior of the DW state (Figs.~2 and 3 in the main text), while the XZ-type MPDW state reproduces the diffusive behavior of the SH state (Figs.~2, 3, and 4 in the main text).  We here choose the system size $L=32$ as examples. 
    }
    \label{fig:Sfig_spin_component}
\end{figure}
\section{II. Properties of the initial states}\label{chapter:Pot}
\subsection{Definitions of the initial states}\label{section:Dot}
Two initial states have been achieved in recent experiments: the DW~\cite{10.1126/science.abk2397,Scheie2021,rosenberg2024dynamics} and spin helix (SH) states~\cite{jepsen2020spin,PhysRevX.11.041054,jepsen2022long}. The DW initial state consists of half-chain spin-up and half-chain spin-down configurations
\begin{equation}
|\varphi_{\mathrm{DW}}\rangle = |\uparrow\rangle^{\otimes L/2} \otimes |\downarrow\rangle^{\otimes L/2}.
\end{equation}
For the SH state, the local wavefunction at site $i$ reads~\cite{jepsen2020spin} 
\begin{equation}
|\varphi_{\mathrm{SH}}\rangle_i = \cos \left( Q i/2 + \theta/2\right)|\uparrow\rangle_i - \sin \left( Q i/2 + \theta/2\right)|\downarrow\rangle_i,
\end{equation}
with wavevector $Q=2\pi/\lambda$ and initial phase $\theta$. 
Hence, the global initial state is the direct product of these local ones $|\varphi_{\mathrm{SH}}\rangle = \bigotimes_i |\varphi_{\mathrm{SH}}\rangle_i$ with the winding pattern in the  $S_x$-$S_z$ plane. Note here that the $z$ component of the local spin exhibits a cosine spatial profile. 

To establish a unified description of the far-from-equilibrium dynamics for the DW and SH states, we introduce the MPDW state with wavevector $Q$ and phase $\theta$. The state's configuration is determined by the amplitude ratio $\eta = |\langle \hat S^x \rangle/\langle\hat S^z\rangle|$ and phase difference $\phi$. 
The local spin state at site $i$ is then defined as
\begin{equation}
|\varphi_{\mathrm{MPDW}}\rangle_i = 
\begin{cases}
A|\uparrow\rangle_i + B|\downarrow\rangle_i, & \cos(Qi+\theta) > 0 ~\text{and}~ \cos(Qi + \theta+\phi) > 0 \\
A|\uparrow\rangle_i - B|\downarrow\rangle_i, & \cos(Qi+\theta) > 0 ~\text{and}~ \cos(Qi + \theta+\phi) < 0 \\
A|\downarrow\rangle_i + B|\uparrow\rangle_i, & \cos(Qi+\theta) < 0 ~\text{and}~ \cos(Qi + \theta+\phi) > 0 \\
A|\downarrow\rangle_i - B|\uparrow\rangle_i, & \cos(Qi+\theta) < 0 ~\text{and}~ \cos(Qi + \theta+\phi) < 0.
\end{cases}
\label{eq:MPDW}
\end{equation}
Here, $k = \sqrt{\eta^2 + 1}$, $A = \sqrt{(k + 1)/(2k)}$, and $B = \sqrt{(k - 1)/(2k)}$. 
The global initial state is given by the direct product of the local ones $|\varphi_{\mathrm{MPDW}}\rangle = \bigotimes_i |\varphi_{\mathrm{MPDW}}\rangle_i$. 

To model the DW state, we employ the Ising-type MPDW initial state with $\eta = 0$, which suppresses the $x$ component of local spins entirely and generates a MPDW configuration only in the $z$ component,
\begin{equation}
|\varphi_{\mathrm{Ising}}\rangle_i = 
\begin{cases}
|\uparrow\rangle_i, & \cos(Qi +\theta) > 0 \\
|\downarrow\rangle_i, & \cos(Qi +\theta) < 0.
\end{cases}
\end{equation}
For the SH state, we employ the XZ-type MPDW state with $\eta = 1$ and $\phi = \pi/2$, exhibiting equal amplitudes in both $x$ and $z$ components of the local spins,
\begin{equation}
|\varphi_{\mathrm{XZ}}\rangle_i = 
\begin{cases}
\sqrt{\frac{\sqrt{2}+1}{2\sqrt{2}}}|\uparrow\rangle_i + \sqrt{\frac{\sqrt{2}-1}{2\sqrt{2}}}|\downarrow\rangle_i, & \cos(Qi+\theta)>0 ~\text{and}~ \cos(Qi+\theta+\frac{\pi}{2})>0 \\
\sqrt{\frac{\sqrt{2}+1}{2\sqrt{2}}}|\uparrow\rangle_i - \sqrt{\frac{\sqrt{2}-1}{2\sqrt{2}}}|\downarrow\rangle_i, & \cos(Qi+\theta)>0 ~\text{and}~ \cos(Qi+\theta+\frac{\pi}{2})<0 \\
\sqrt{\frac{\sqrt{2}+1}{2\sqrt{2}}}|\downarrow\rangle_i + \sqrt{\frac{\sqrt{2}-1}{2\sqrt{2}}}|\uparrow\rangle_i, & \cos(Qi+\theta)<0 ~\text{and}~ \cos(Qi+\theta+\frac{\pi}{2})>0 \\
\sqrt{\frac{\sqrt{2}+1}{2\sqrt{2}}}|\downarrow\rangle_i - \sqrt{\frac{\sqrt{2}-1}{2\sqrt{2}}}|\uparrow\rangle_i, & \cos(Qi+\theta)<0 ~\text{and}~ \cos(Qi+\theta+\frac{\pi}{2})<0.
\end{cases}
\end{equation}

Fig.~\ref{fig:Sfig_spin_component}(a-d) respectively shows the DW, SH, Ising-type MPDW, and XZ-type MPDW initial states, used in the main text. By examining the scaling exponents of the first few cumulants employing both polarization transfer and contrast, we investigate transport dynamics of the isotropic Heisenberg spin chain initialized from these MPDW states. Our analysis reveals that the MPDW state reproduces the distinct scaling characteristic of both the SH state ($\nu_1\approx 0.45$, diffusive) and the DW state ($\nu_1\approx 0.65$, superdiffusive) in their respective regimes, as show in Figs.~2, 3, and 4 of the main text.

\begin{figure}[t!]
    \centering
    \includegraphics[width=1.0\textwidth]{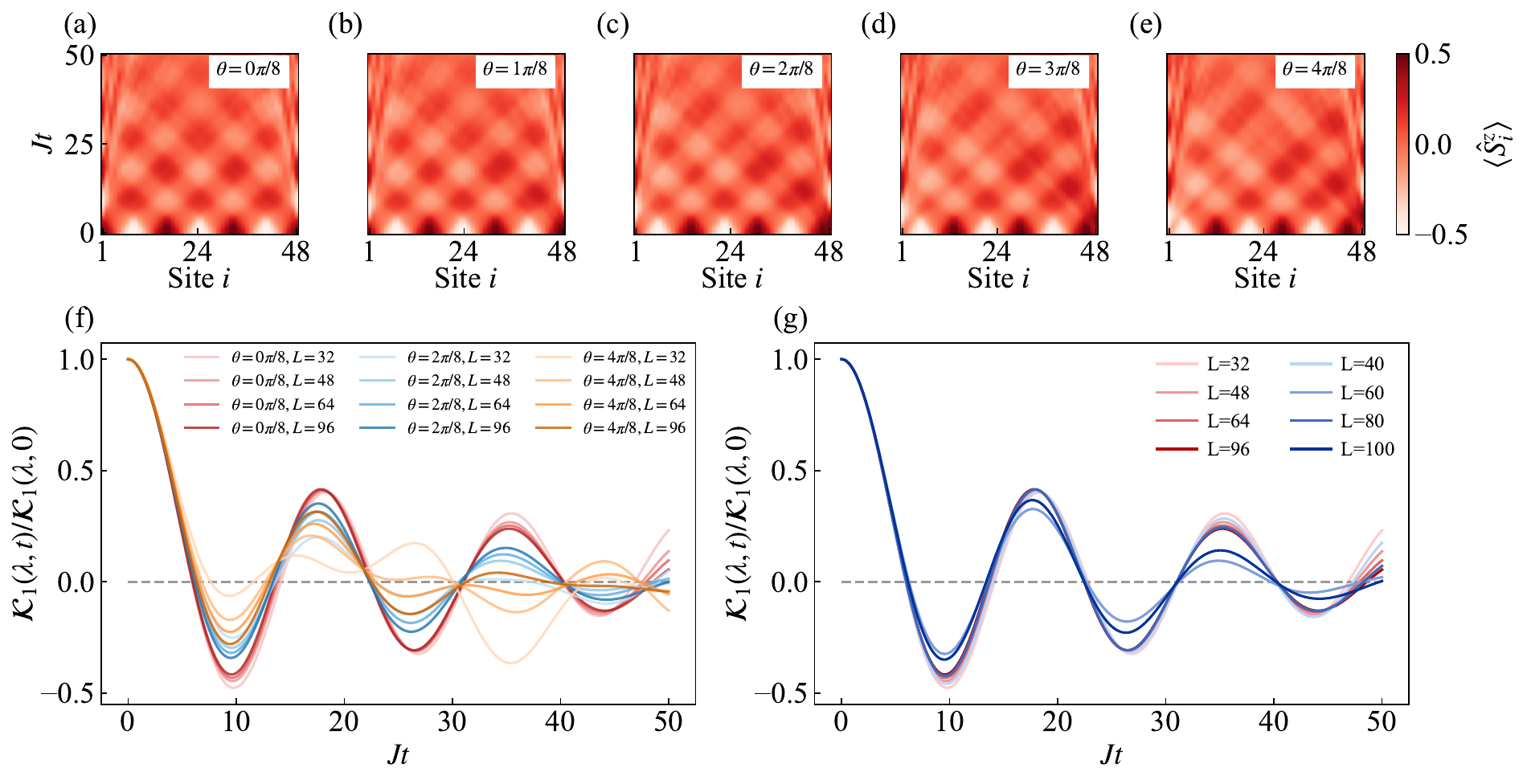}
    \caption{%
        Phase and size dependencies of the normalized first-order contrast $\mathcal{K}_1(\lambda, t)/\mathcal{K}_1(\lambda, 0)$ for the SH states with $\lambda=16$ in the XX model. %
        (a-e) Real-space time evolution of $\langle \hat{S}^z_i\rangle$ in a system $L=48$ with initial phases $\theta = 0$, $\pi/8$, $\pi/4$, $3\pi/8$, and $\pi/2$. 
        (f) Initial phase dependence of the first-order contrast cumulant. Colors from light to dark represent $L = 32$, 48, 64, and 96, where the $\theta = 0$ cases (red lines) exhibit optimal robustness. 
        (g) System-size dependence of the first-order contrast cumulant with $\lambda=16$ and $\theta=0$. 
        Light-to-dark red lines denote the systems with $L = 32$, 48, 64, and 96, corresponding to integer multiples of the wavelength $\lambda$, while light-to-dark blue lines represent the systems with $L = 40$, 60, 80, and 100, 
        representing non-integer multiples of $\lambda$.
        Here, we choose $\chi=300$.
    }
    \label{fig:Sfig_N_theta_conv}
\end{figure}

\subsection{Parameter sensitivity of multi-periodic states}\label{section:Pso}
Scaling dynamics of the contrast cumulant is generally independent of the parameters of the multi-periodic state, such as initial phase $\theta$ and wavelength $\lambda$. However, parameters significantly influences the long-time dynamics in finite chains due to its association with boundary effects, playing an important role in the scaling extraction.
To study the influence of the parameters on the transport dynamics, we perform systematic phase-sweep analysis using the XX model~\cite{jepsen2020spin},
\begin{equation}
\hat{H}_{\text{XX}} = -J \sum_i \left( \hat{S}^x_i \hat{S}^x_{i+1} + \hat{S}^y_i \hat{S}^y_{i+1} \right).
\label{eq:H_XXZ}
\end{equation}
It is predicted that the first-order contrast cumulant follows Bessel-function dynamics, providing distinct oscillation signatures for parameter sensitivity analysis~\cite{PhysRevA.110.023313,PhysRevA.106.043306,PhysRevB.108.075135}.

Using the SH initial state at fixed $\lambda=16$ as an example, we reveal how the transport dynamics depends on both system size $L$ and 
initial phase $\theta$.
Fig.~\ref{fig:Sfig_N_theta_conv}(a-e) show the time evolution of spin distributions for the systems with $L = n_p \lambda$ ($n_p=3$, number of periods), with $\theta$ varying from $0$ to $\pi/2$.
We find that boundary effects are minimized relative to $\theta$ when the $z$-component spin at the boundaries is maximized ($\theta=0$), as demonstrated in Fig.~\ref{fig:Sfig_N_theta_conv}(f). Furthermore, optimal convergence with respect to $L$ occurs when the system size satisfies the integer-wavelength condition $L = n_p \lambda$ ($n_p \in \mathbb{Z}$), as shown in Fig.~\ref{fig:Sfig_N_theta_conv}(g).
Although longer-wavelength initial states reduce boundary effects, they require prohibitively long evolution time to observe pronounced relaxation dynamics~\cite{jepsen2020spin,PhysRevB.104.L081410}.
To access scaling behaviors of higher-order cumulants within experimentally feasible timescales, we therefore employ shorter-wavelength initial states for studying the transport dynamics.
Consequently, balancing these factors---minimizing boundary effects ($\theta=0$, $L = n_p \lambda$), and enabling feasible observation time (shorter $\lambda$)---our main-text simulations mainly use a system size $L=96$ and multi-periodic initial states with phase $\theta=0$ and integer wavelengths ($\lambda=8,~12,~16,~24,~32$).

\begin{figure}[t!]
    \centering \includegraphics[width=0.9\textwidth]{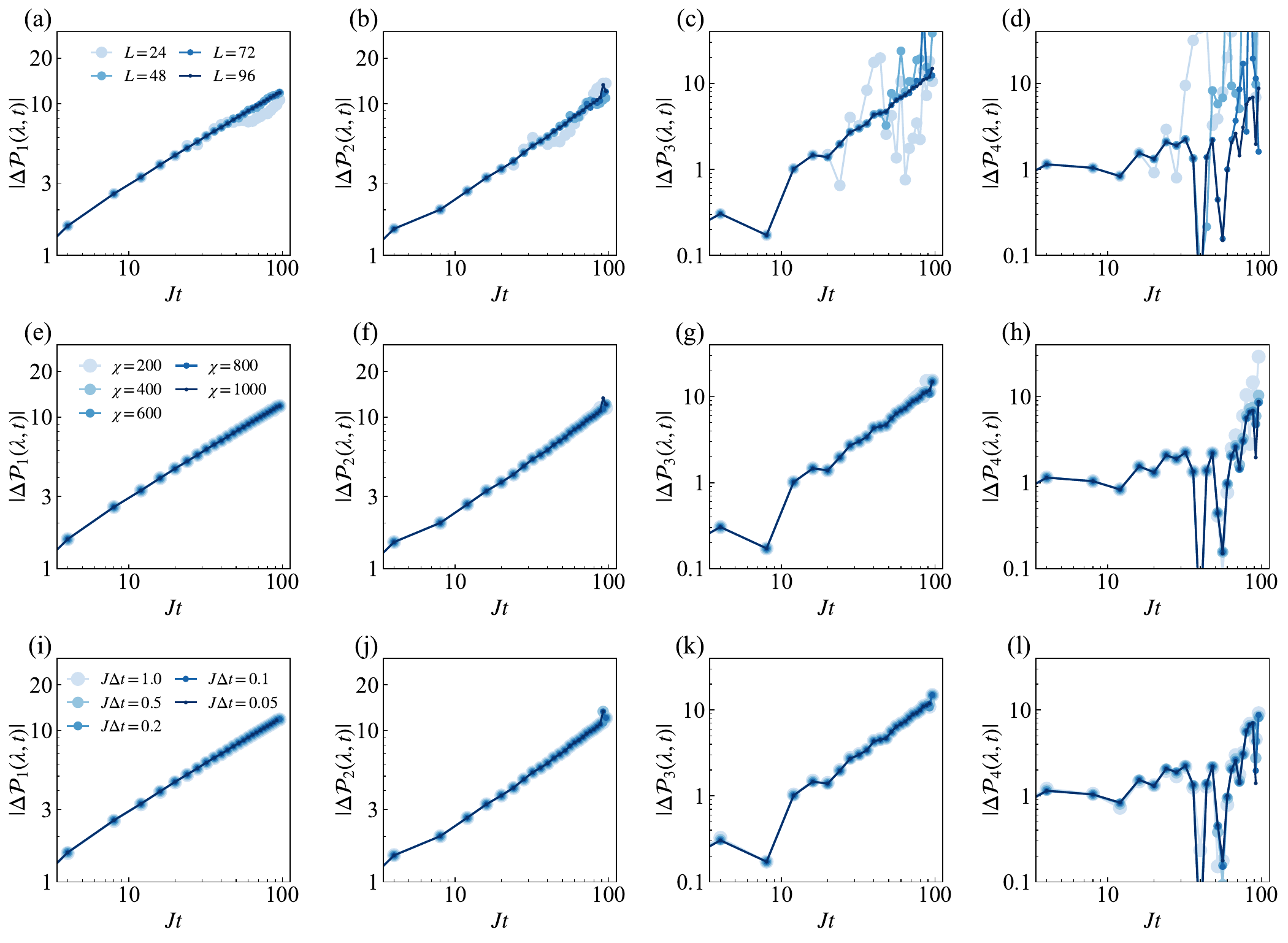}
    \caption{%
    Convergence analysis of spin polarization transfer cumulants $|\Delta{\mathcal P}_n(\lambda, t)|$ for the DW initial states. %
    Left to right columns respectively correspond to the first- to fourth-order cumulants. %
    (a-d) Cumulants of polarization transfer for different system sizes $L = 24$, 48, 72, and 96 (light to dark) with MPS bond dimension $\chi=1000$ and time step $J \Delta t=0.1$. %
    (e-h) Cumulants for different bond dimensions $\chi = 200$, 400, 600, 800, and 1000 (light to dark) with $L=96$ and $J \Delta t=0.1$. %
    (i-l) Cumulants for different time steps $J \Delta t = 1.0$, 0.5, 0.2, 0.1, and 0.05 (light to dark) with $\chi=1000$ and $L=96$. %
    }
    \label{fig:Sfig_Pt_conv}
\end{figure}

\section{III. Convergence tests in the far-from-equilibrium dynamics }\label{chapter:Ctd}
Characterizing far-from-equilibrium signatures requires rigorous convergence tests. This enables reliable extraction of dynamical scaling exponents in the long-time dynamics. In this section, we investigate quantum cumulant dynamics of spin polarization transfer and contrast to ensure full convergence. In the main text, we employ open boundary conditions with system size $L=96$, bond dimensions $\chi = 1000$ (spin polarization transfer) and $500$ (contrast), and time step $J \Delta t = 0.1$. These parameter values are validated by convergence tests presented below.

\begin{figure}[t!]
    \centering
    \includegraphics[width=0.9\textwidth]{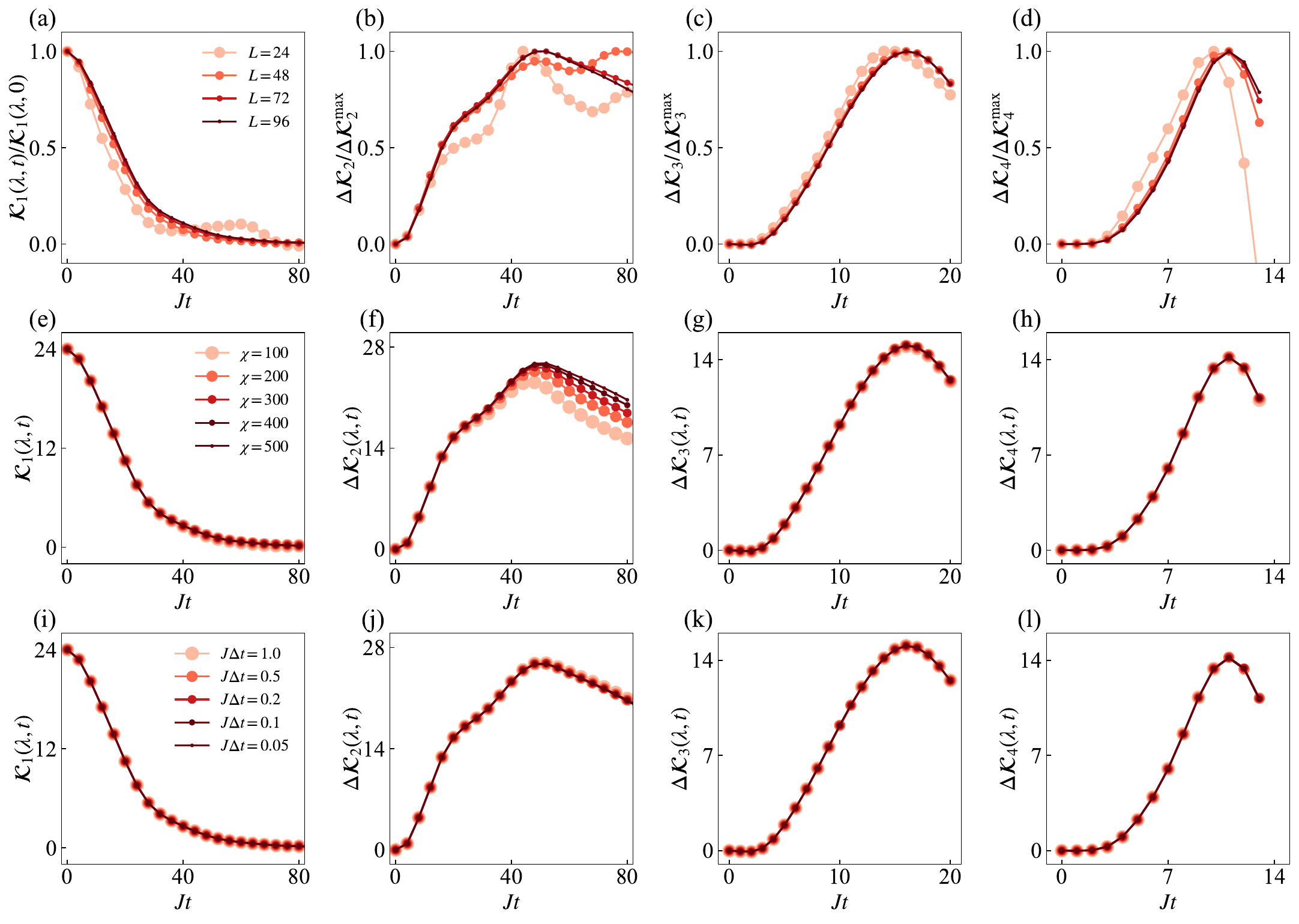}
    \caption{Convergence analysis of contrast cumulants for the SH state with $\lambda=12$. %
        Left to right columns respectively correspond to first- to fourth-order cumulants. %
        (a-d) System-size analysis: $L = 24$, 48, 72, and 96 (light to dark) with $\chi=500$ and $J \Delta t=0.1$. %
        (e-h) Bond-dimension convergence: $\chi = 100$, 200, 300, 400, and 500 (light to dark) with $L=96$ and $J \Delta t=0.1$. %
        (i-l) Time-step convergence: $J \Delta t = 1.0$, 0.5, 0.2, 0.1, and 0.05 (light to dark) with $\chi=500$ and $L=96$. %
    }
    \label{fig:Sfig_Ct_conv}
\end{figure}

\subsection{Cumulant dynamics of spin polarization transfer}\label{section:Cos}

In the part, we investigate the quantum cumulant dynamics in the framework of spin polarization transfer, and verify the convergence of the cumulants against system size, MPS bond dimension, and time step. We find that the parameters used in the main text, {\it i.e.} lattice size $L=96$, bond dimension $\chi=1000$, and time step $J\Delta t=0.1$, yield converged cumulant dynamics of spin polarization transfer. 

Without loss of generality, we focus on the single-periodic Ising-type DW initial state with $\lambda =L$.
Fig.~\ref{fig:Sfig_Pt_conv} presents a systematic convergence analysis for the first four cumulants $|\Delta\mathcal{P}_{n}(\lambda, t)|$ of spin polarization transfer, examining dependencies on (a-d) system size $L$, (e-h) MPS bond dimension $\chi$, and (i-l) time step $\Delta t$. 
The first row (a-d) assesses system size convergence. Although boundary reflections emerge at $Jt \approx L/2$ for the single-periodic DW states~\cite{rosenberg2024dynamics,10.1126/science.abk2397}, simulations extend to $Jt \approx L$ to access longer timescales. We observe excellent convergence for the first- and second-order cumulants [$Jt \approx L$, (a)(b)]. In contrast, the third- and fourth-order ones exhibit stronger boundary effects at earlier time ($Jt \approx L/2$), particularly in smaller systems (c)(d). We find that system size with $L=96$ is sufficient to mitigate boundary artifacts and probe emergent long-time scaling laws.

Convergence tests on bond dimensions are shown in the second row [Fig.~\ref{fig:Sfig_Pt_conv}(e-h)]. We reveal that the first two cumulant orders are consistent even at $\chi=200$, while higher-order cumulants require $\chi=800$. We therefore employ $\chi=1000$ to ensure accuracy for all higher-order cumulants in the timescale studied here. Time step convergence [third row, (i-l)] shows increasingly stringent precision requirements for higher-order cumulants, with $J\Delta t=0.1$ proving reliable for capturing the long-time dynamics.
These comprehensive convergence tests establish that the parameters---system size $L=96$, time step $J\Delta t=0.1$, and bond dimension $\chi=1000$---reliably capture many-body dynamics before boundary effects occur ($Jt<L/2$), as discussed in Figs. 2 and 3 of the main text.
Finally, for the multi-periodic states, our analysis indicates that the scaling laws are valid within certain time windows bounded by boundary reflections: $Jt \lesssim \lambda/2$ for the Ising-type MPDW states and $Jt \lesssim \lambda/4$ for the XZ-type MPDW states, as discussed below in Figs.~\ref{fig:Sfig_Pt_k1_to_k4} and \ref{fig:Sfig_Pt_S_and_Q.pdf}.

\subsection{Cumulant dynamics of contrast analysis }\label{section:Coc}
In the part, we investigate the quantum cumulant dynamics in the framework of contrast, and verify the convergence of cumulants with respect to system size, MPS bond, and time step. We find that the parameters used in the main text, including lattice size $L=96$, bond dimension $\chi=500$, and time step $J\Delta t=0.1$, yield converged contrast cumulant dynamics. 

We take the SH initial state as an example. As shown in Fig.~\ref{fig:Sfig_Ct_conv}(a-d), systems of $L>72$ are sufficient to examine scaling dynamics while simultaneously suppressing boundary-induced disturbances: $\mathcal{K}_{1}$ remains well converged up to the timescale where it decays to 0.4, whereas $\mathcal{K}_{2,3,4}$ achieve convergence at their respective maxima.
On this basis, our scaling analysis of cumulant dynamics relies on numerical data in the regime from the initial state to the threshold time $\tau$: the decay time to 0.4 for $\mathcal {K}_{1}$ and the attainment of the first peak values for $\mathcal {K}_{2,3,4}$. Within this time window, universal scaling behavior is reliably captured, as shown in Fig.~4 of the main text and Figs.~\ref{fig:Sfig_SH_rescale_scaling} and ~\ref{fig:Sfig_XZ_rescale_scaling} below. 

The dynamics of contrast cumulant exhibit strong bond-dimension dependence in MPS simulations. As anticipated, higher-order cumulants and longer evolution time necessitate larger bond dimension $\chi$. We confirm that $\chi = 500$ is sufficient for precise extraction of the threshold time $\tau$ [Fig.~\ref{fig:Sfig_Ct_conv}(e-h)] and scaling exponents [Figs.~\ref{fig:Sfig_SH_rescale_scaling} and \ref{fig:Sfig_XZ_rescale_scaling}]. The apparent insensitivity of $\mathcal{K}_{3,4}$ to $\chi$ arises from the relatively short evolution times considered. The dynamics beyond $\tau$ is similarly expected to exhibit identical scaling behavior, but its resolution requires larger $L$ and $\chi$ for precise characterization, which lies beyond the scope of the present work.
In addition, time step variations in Fig.~\ref{fig:Sfig_Ct_conv}(i-l) show minimal impact on our dynamical results. We therefore choose $J \Delta t=0.1$ for all contrast cumulant calculations.

\begin{figure}[t!]
    \centering    \includegraphics[width=0.8\textwidth]{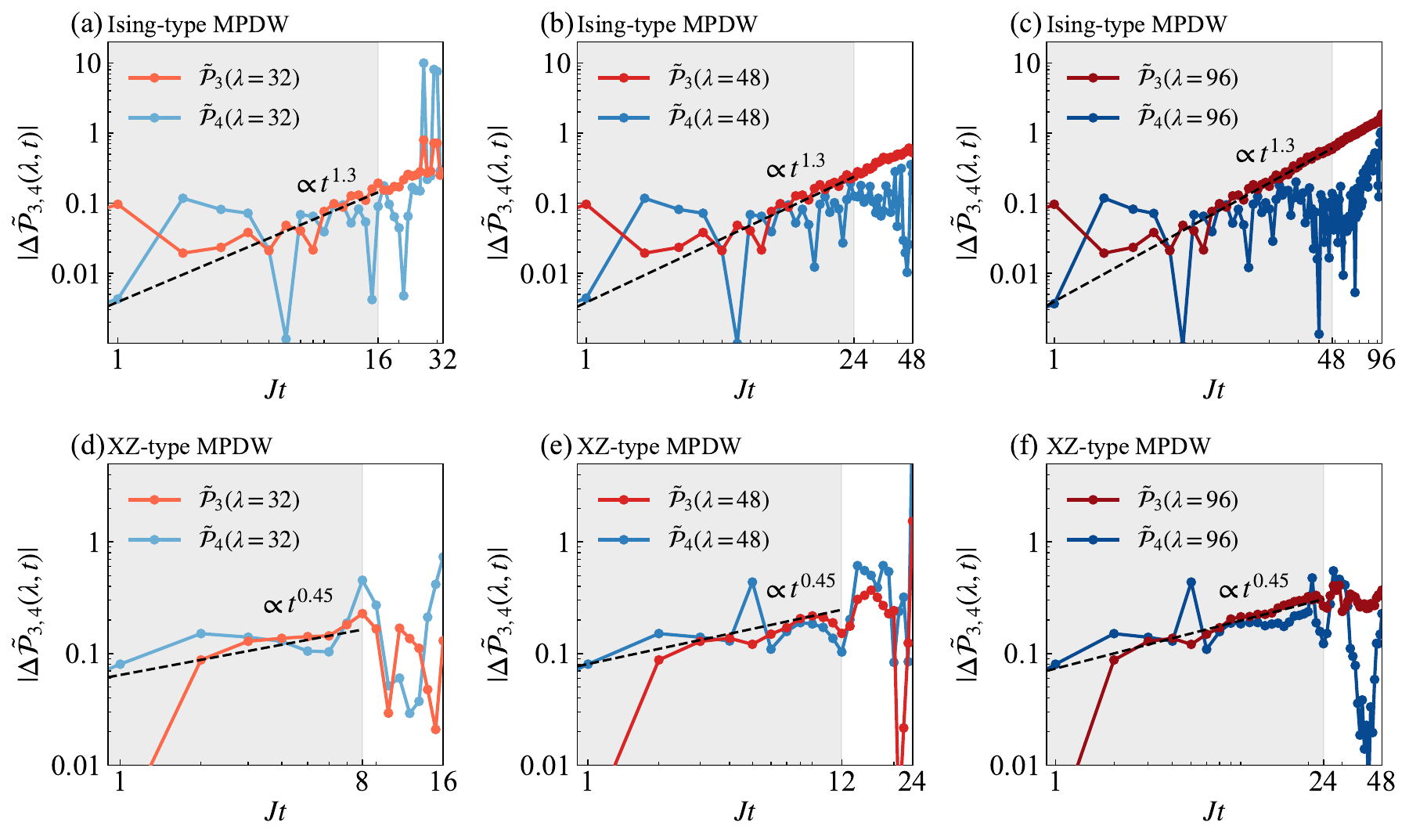}
    \caption{Averaged polarization transfer cumulants $|\Delta{\tilde{\mathcal P}}_{3,4}(\lambda, t)|$ for the Ising-type (top row) and XZ-type MPDW (bottom row) states. The columns from left to right corresponds to the initial states with  wavelengths $\lambda=32$, $48$, and $96$. Black dashed lines guide power-law relation $|\Delta{\tilde{\mathcal P}}_{3,4}(\lambda,t)| \propto t^{\nu_{3,4}}$ with $\nu_{3, 4}\approx1.3$ for the Ising-type MPDW states, and $\nu_{3, 4}\approx0.45$ for the XZ-type MPDW states. Gray shaded regions mark the boundary-free timescales for each initial state with wavelength $\lambda$, $\it i.e.$ $Jt \lesssim \lambda/2$ for the Ising-type states, and $Jt \lesssim \lambda/4$ for the XZ-type MPDW states, respectively. Here, system size $L=96$, bond dimension $\chi=1000$, and time step $J\Delta t=0.1$. 
    }
    \label{fig:Sfig_Pt_k1_to_k4}
\end{figure}
\begin{figure}[h!]
    \centering    \includegraphics[width=0.8\textwidth]{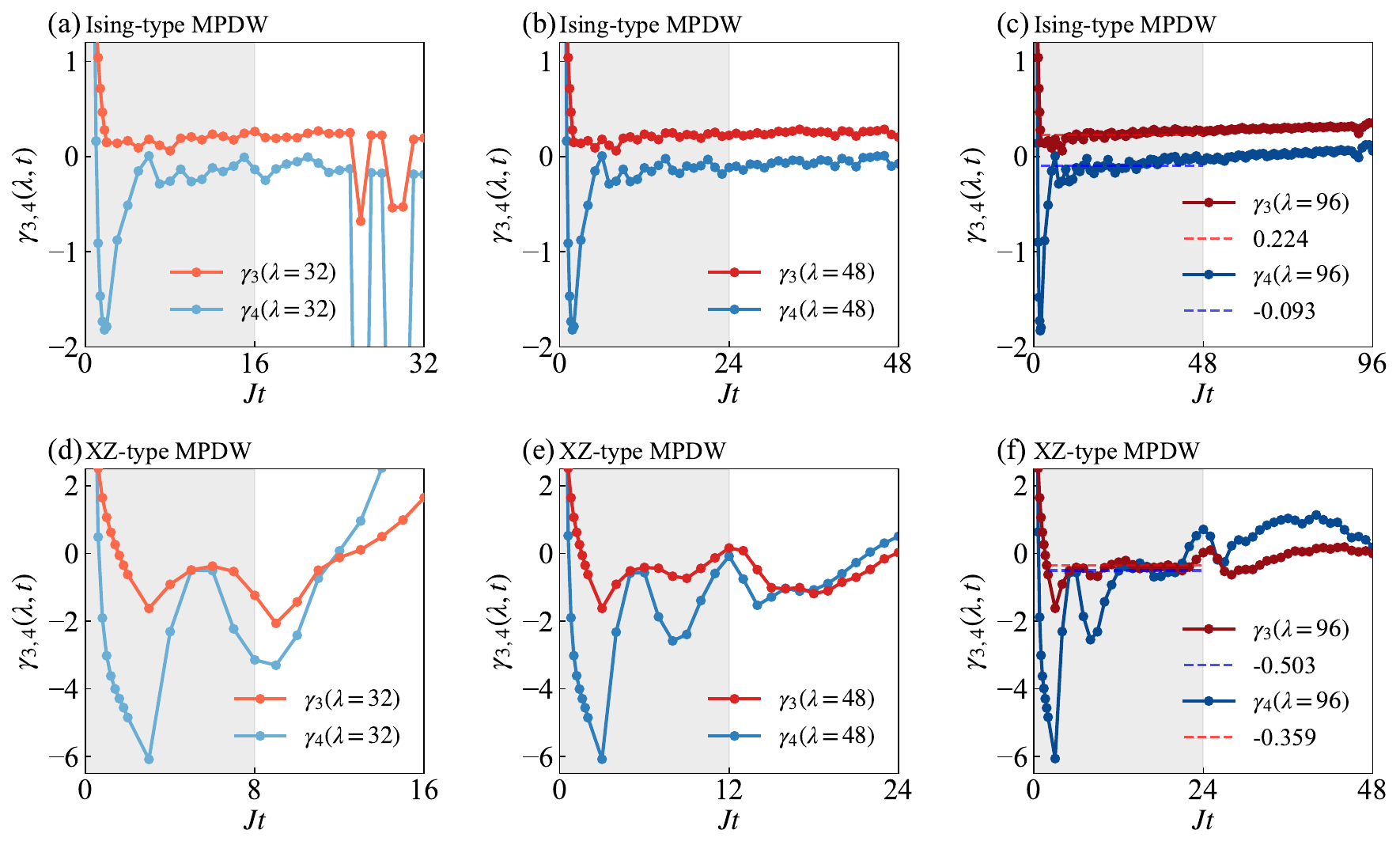}
    \caption{Standardized cumulants $\gamma_{3,4}(\lambda, t)$ for the Ising-type (top row) and the XZ-type MPDW (bottom row) states. The red and blue curves denote skewness $\gamma_3(\lambda, t)$ and excess kurtosis $\gamma_4(\lambda, t)$, respectively. 
        Other parameters and legends are consistent with Fig.~\ref{fig:Sfig_Pt_k1_to_k4}. %
    }
    \label{fig:Sfig_Pt_S_and_Q.pdf}
\end{figure}

\section{IV. Additional data of polarization transfer and contrast cumulants}\label{chapter:Ado}

\subsection{Spin polarization transfer scaling for Ising-type and XZ-type MPDW states}\label{section:Spt}
In this part, we provide more information for the third- and fourth-order cumulants of the spin polarization transfer, where the system is initialized from the Ising-type and XZ-type MPDW states. To ensure consistent quantification of polarization transfer across periodic structures, we define the averaged transfer cumulants
\begin{equation}
\Delta{\tilde{\mathcal P}}_n(\lambda,t) \equiv \Delta{\mathcal P}_n(\lambda,t) / n_c,
\label{eq:averaged_transfer}
\end{equation}
where $n_c$ counts the number of up-down edge points in the initial DW structure, related to the number of periods $n_p\equiv L/\lambda$ by $n_c = 2n_p - 1$ yielding $n_c = 1, 3, 5$ for $n_p = 1, 2, 3$, respectively.

Fig.~\ref{fig:Sfig_Pt_k1_to_k4} presents the third- and fourth-order cumulants of spin polarization transfer for the Ising-type (top row) and XZ-type (bottom row) MPDW states in the system with $L=96$. Without loss of generality, we consider three different cases with (a)(d) $n_p=3$, (b)(e) = 2, and (c)(f) = 1.
Within the boundary-free timescales (shaded regions), higher-order cumulants display identical power-law scaling across different $n_p$. 
Specifically, this scaling persists up to $Jt\lesssim \lambda/2$ for the Ising-type MPDW states and $Jt\lesssim\lambda/4$ for the XZ-type MPDW states.
Beyond these timescales, boundary effects induce pronounced  oscillations in the higher-order cumulants. 
Nevertheless, reliable scaling exponents can be extracted from the finite time window: $\nu_{1,2} \approx 0.65$ and $\nu_{3,4} \approx 1.3$ for the Ising-type states, and $\nu_{1,2,3,4}\approx0.45$ for the XZ-type states. 
These period-independent scaling exponents confirm universal transport behavior in the higher-order cumulants, consistent with the main text conclusions in Fig.~3(a)(b).
Note here that we only consider the cumulants on spin-up sites in Fig.~\ref{fig:Sfig_Pt_k1_to_k4} and Fig.~3(a)(b) in the main text. This site-selective approach suppresses finite-size effects exacerbated by collective behaviors, as demonstrated in Ref.~\cite{wang2025eigenstatethermalizationhypothesiscorrelations}.

\begin{figure}[t!]
    \centering \includegraphics[width=0.8\textwidth]{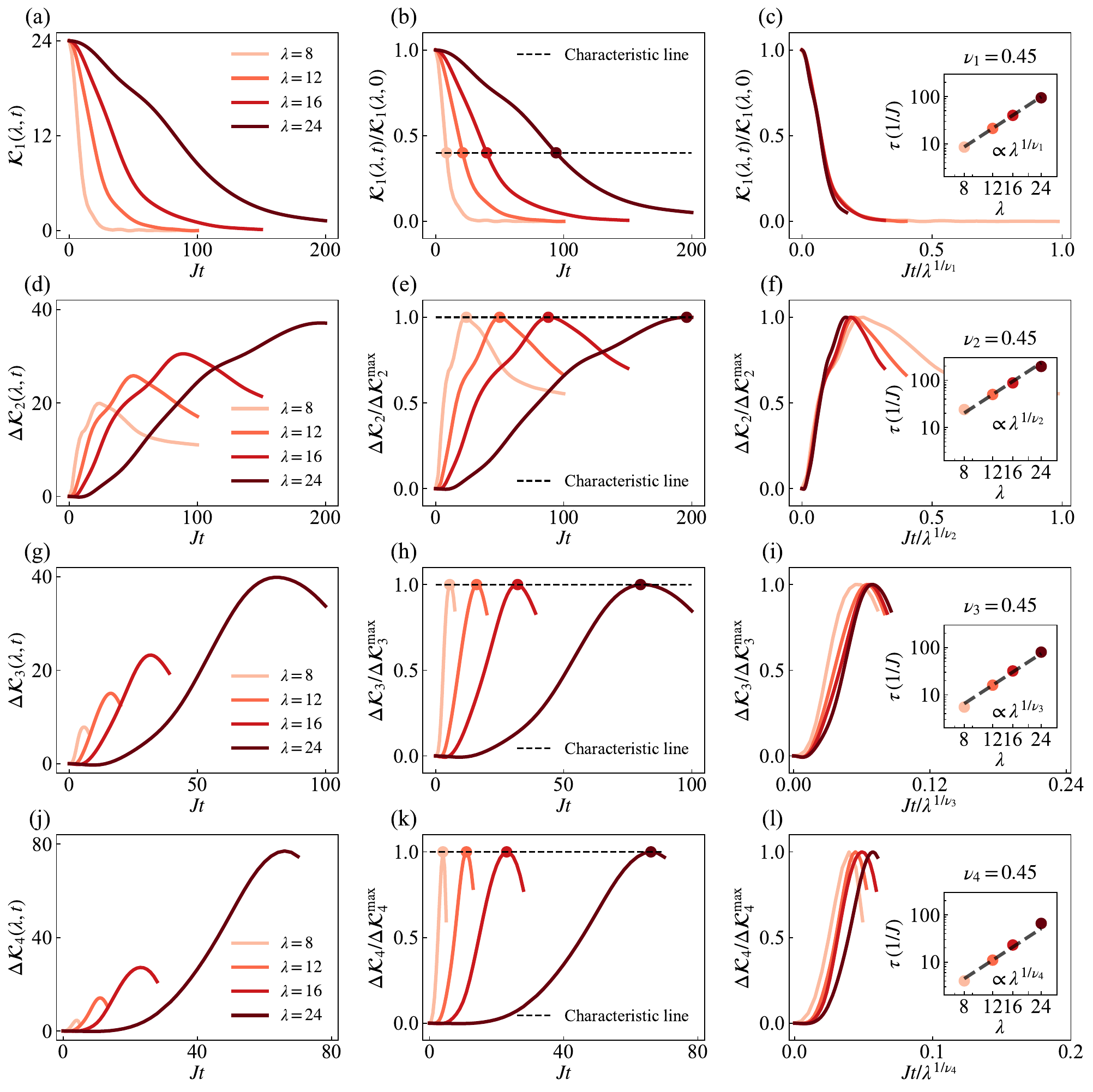}
    \caption{
        Data collapse and scaling analysis of contrast cumulants $\mathcal{K}_n(\lambda, t)$ for the SH states. %
        Rows: First- to fourth-order contrast cumulants (top to bottom). %
        Columns: (a)(d)(g)(j) Original cumulant versus time. (b)(e)(h)(k) Cumulants, rescaled by their initial (first-order) or peak (higher-order) values, are plotted against time, where the dashed lines indicate the characteristic values for threshold time extraction. (c)(f)(i)(l) Rescaled cumulants are plotted against time rescaled by $\lambda^{1/\nu_n}$. Insets: Power-law scaling $\tau \propto \lambda^{1/\nu_n}$ between the wavelength $\lambda$ and threshold time $\tau$, confirming $\nu_{1, 2, 3, 4} \approx 0.45$ for the SH initial state. Here, we choose $L=96$, $\chi=500$, and $J\Delta t=0.1$. %
        }
\label{fig:Sfig_SH_rescale_scaling}
\end{figure}
\begin{figure}[h!]
    \centering
\includegraphics[width=0.8\textwidth]{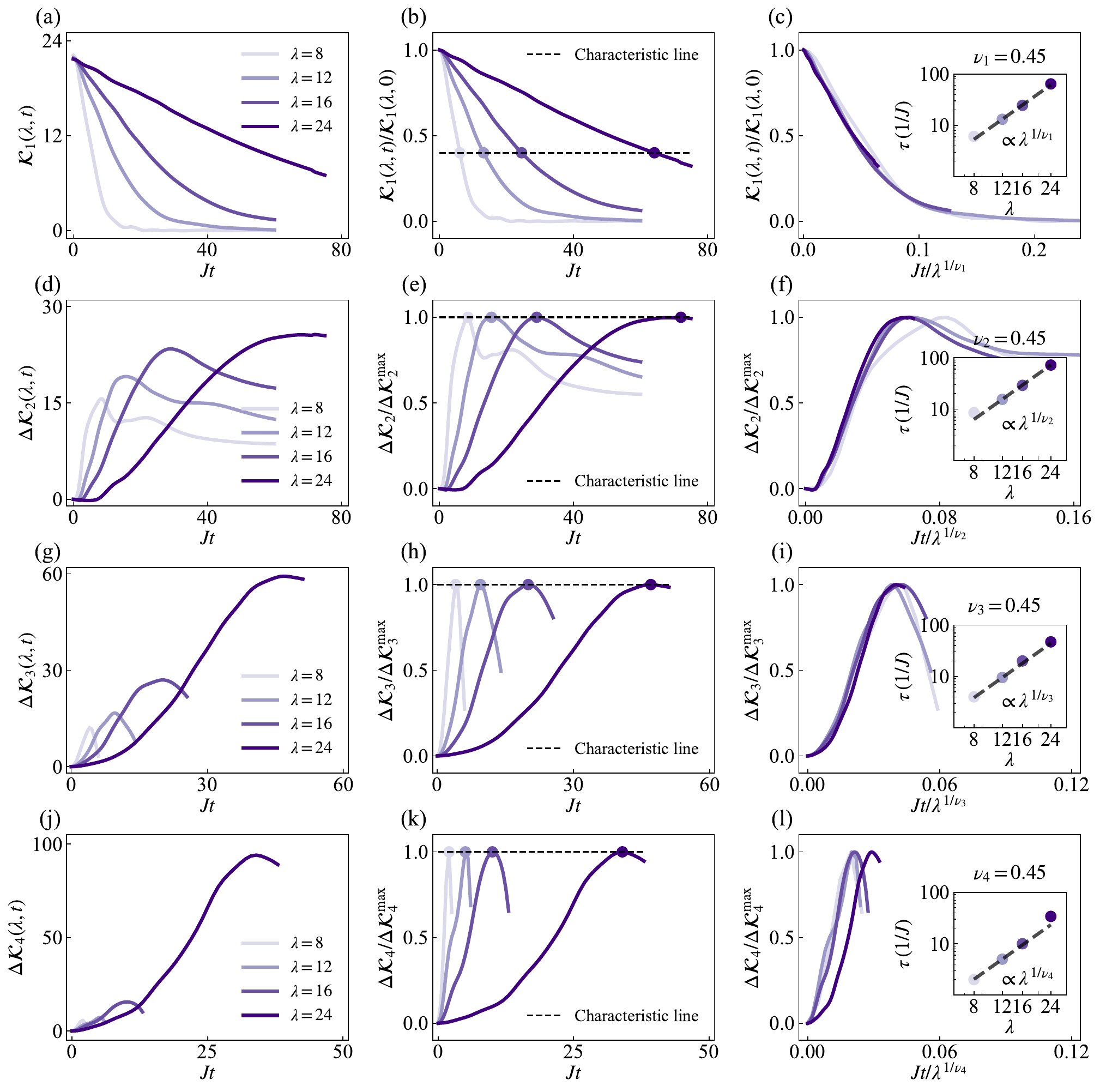}
    \caption{
        Data collapse and scaling analysis of contrast cumulants $\mathcal{K}_n(\lambda, t)$ for the XZ-type MPDW states. Rows: First- to fourth-order contrast cumulants (top to bottom). %
        Columns: (a)(d)(g)(j) Original cumulant versus time, (b)(e)(h)(k) rescaled cumulant versus time, and (c)(f)(i)(l) rescaled cumulants versus rescaled time by scaling exponent $\nu_n$. 
        Format and analysis are identical to Fig.~\ref{fig:Sfig_SH_rescale_scaling}, confirming identical scaling exponents $\nu_{1, 2, 3, 4} \approx 0.45$ for the XZ-type MPDW state. Here, we choose $L=96$, $\chi=500$, and $J\Delta t=0.1$.
    }  \label{fig:Sfig_XZ_rescale_scaling}
\end{figure}

We define the standardized cumulant using averaged transfer cumulant in Eq.~(\ref{eq:averaged_transfer}), which is given by
\begin{equation}
\begin{aligned}
    \gamma_3(\lambda,t) &= \Delta{\tilde{\mathcal P}}_3(\lambda,t) / \Delta{\tilde{\mathcal P}}_2^{3/2}(\lambda,t),\\
    \gamma_4(\lambda,t) &=\Delta{\tilde{\mathcal P}}_4(\lambda,t)/ \Delta{\tilde{\mathcal P}}_2^{2}(\lambda,t).
\end{aligned}
\label{eq:S_and_Q}
\end{equation}
Fig.~\ref{fig:Sfig_Pt_S_and_Q.pdf} presents the dynamics of standardized third- [$\gamma_3(\lambda,t)$, skewness] and fourth-order [$\gamma_4(\lambda,t)$, excess kurtosis] cumulants for the Ising-type and XZ-type MPDW initial states. Within boundary-free shaded regions, both standardized cumulants show consistent and stable values across initial states. Boundary effects appear at $Jt>\lambda/2$ for the Ising-type MPDW states and $Jt>\lambda/4$ for the XZ-type MPDW states, manifesting significantly earlier and with greater amplitude for shorter wavelengths ($\lambda=32$) compared to longer ones ($\lambda=96$).
Analysis of the standardized cumulants (Fig.~\ref{fig:Sfig_Pt_S_and_Q.pdf}) further confirms that the boundary-free regions yield reliable results, consistent with the stable power-law scaling shown in Fig.~\ref{fig:Sfig_Pt_k1_to_k4}.
These results across wavelengths confirm that the late-time increases in skewness $\gamma_3(\lambda,t)$ and excess kurtosis $\gamma_4(\lambda,t)$---visible in Fig.~3(c)(d) of the main text---are finite-size effects caused by domain-wall boundaries.

\subsection{Contrast scaling for SH and XZ-type MPDW states}\label{section:Csi}
In this part, we provide supplementary details on the power-law scaling of the contrast functions for the SH and XZ-type MPDW states, as shown in Fig.~4 in the main text. Generally, distinct cumulants of conserved quantities capture different nonlocal multipoint correlations, each exhibiting unique dynamical timescales towards equilibrium in the time evolution. However, these nonlocal correlations are expected to follow certain scaling laws in the long-time dynamics.

We analysis the wavelength dependence of the threshold time $\tau$ for the first few cumulants of contrast $\mathcal{K}_n(\lambda, t)$ for both SH (Fig.~\ref{fig:Sfig_SH_rescale_scaling}) and XZ-type MPDW states (Fig.~\ref{fig:Sfig_XZ_rescale_scaling}). As shown in Figs.~\ref{fig:Sfig_SH_rescale_scaling}(a)(d)(g)(j) and \ref{fig:Sfig_XZ_rescale_scaling}(a)(d)(g)(j), the system exhibits distinct evolution timescales across various wavelengths for different cumulants. To quantify the associated scaling exponents, we define wavelength-dependent threshold time $\tau$. As shown by the dashed lines in Figs.~\ref{fig:Sfig_SH_rescale_scaling}(b)(e)(h)(k) and \ref{fig:Sfig_XZ_rescale_scaling}(b)(e)(h)(k), $\tau$
for $n=1$ is defined as the decay time from 1.0 to 0.4, whereas for $n=2$, 3, and 4, $\tau$ is defined as the time at which the first maxima occurs. Here, all the corresponding cumulants are rescaled by their respective maxima.
The third columns of Figs.~\ref{fig:Sfig_SH_rescale_scaling} and \ref{fig:Sfig_XZ_rescale_scaling} demonstrate a universal temporal scaling collapse of contrast ${\mathcal K}_{1,2,3,4}$ when rescaled by $\lambda^{1/\nu_n}$. This scaling is reconfirmed by the power-law behavior $\tau \propto \lambda^{1/\nu_n}$ for the threshold time $\tau$ [insets of Figs.~\ref{fig:Sfig_SH_rescale_scaling}(c)(f)(i)(l) and \ref{fig:Sfig_XZ_rescale_scaling}(c)(f)(i)(l)].  
These results provide complementary data to Fig.~4 of the main text and establish identical diffusive scaling, characterized by the universal exponent $\nu_{1,2,3,4} \approx 0.45$, for the first four cumulants in both SH and XZ-type MPDW states.

\section{V. Dynamical scaling of subsystem fluctuations}\label{chapter:Smf}
\begin{figure}[t!]
    \centering
\includegraphics[width=0.9\textwidth]{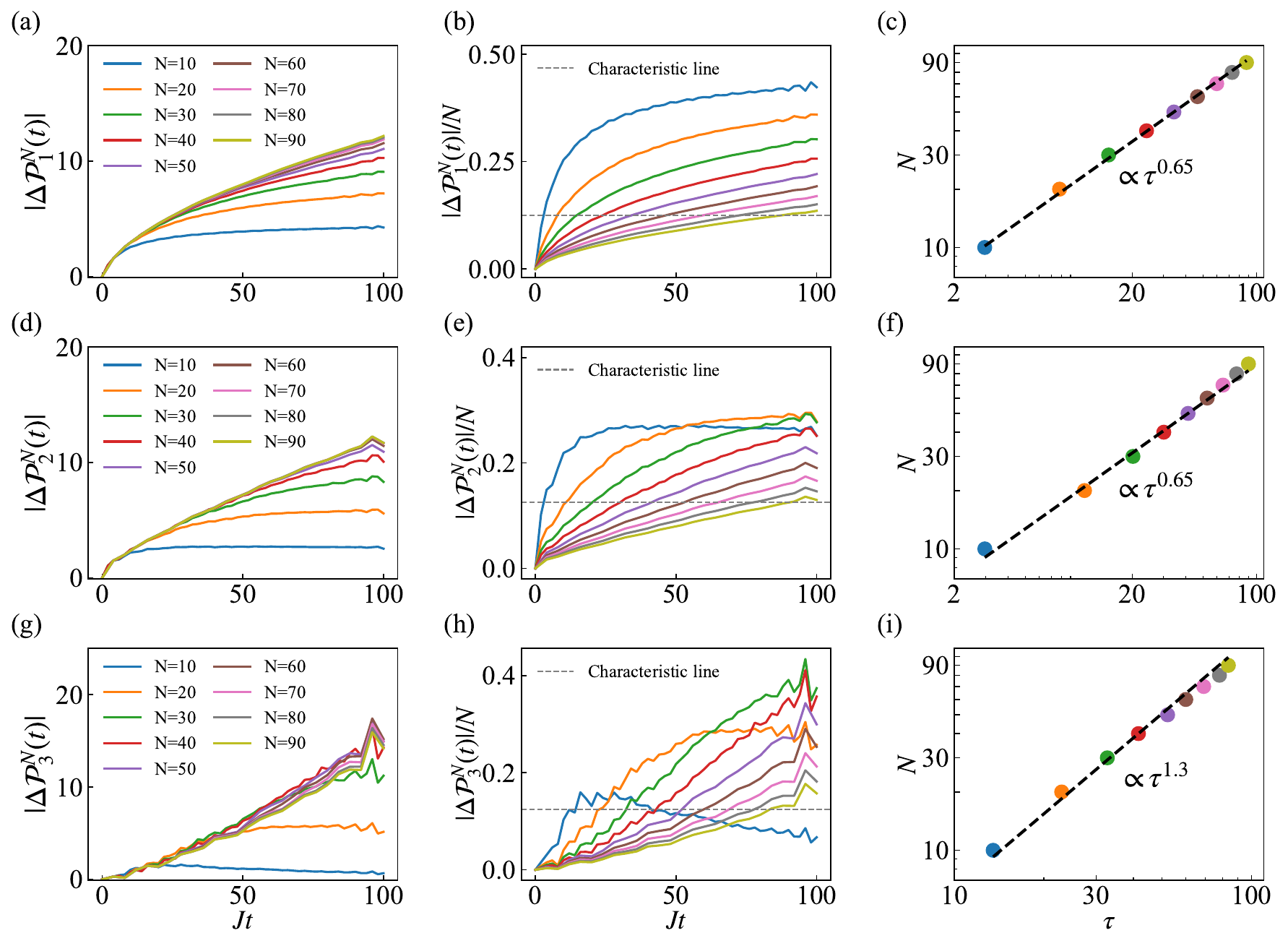}
    \caption{Dynamics of local observation values $|\Delta{\mathcal P}^N_1 (t)|$ and their subsystem fluctuations $|\Delta{\mathcal P}^N_{2,3} (t)|$ for the DW initial state. Rows from top to bottom display the first-, second-, and third-order values, respectively. (a)(d)(g) show raw cumulants from Eq.~(\ref{eq:subsystem_cumulants}) for the subsystem size $N=10$ to $90$. (b)(e)(h) show the size-normalized cumulants, where the dashed lines denote the characteristic value 0.15. (c)(f)(i) show the threshold time $\tau$ as a function of subsystem size $N$, which exhibits universal power-law scalings $N \propto \tau^{\nu_n}$ with the black dashed lines indicating power-law guides. The lattice size $L=96$, bond dimension $\chi=300$, and time step $J\Delta t=0.1$.
    } \label{fig:Sfig_subsystem_scaling}
\end{figure}

In this section, we  investigate the scaling dynamics of subsystem fluctuations in the Heisenberg spin chain. This quantify is calculated from the spin-spin correlations, which are obtained by computing the subsystem cumulants
\begin{equation}
    \mathcal{P}_n^N(t) = \sum_{i_1, \cdots, i_n=1}^{N} \langle \hat{\mathcal{S}}_{i_1}^z \hat{\mathcal{S}}_{i_2}^z \cdots \hat{\mathcal{S}}_{i_n}^z \rangle_c,
    \label{eq:subsystem_cumulants}
\end{equation}
where $N$ denotes the size of the subsystem and $\Delta{\mathcal P}_n^N(t) \equiv {\mathcal P}_n^N(t)- {\mathcal P}_n^N(0)$ represents the $n$th-order cumulants of transferred spin polarization on the subsystem. Specifically, $\Delta{\mathcal P}^N_1 (t)$ denotes the local observation values, and $\Delta{\mathcal P}^N_{n>1} (t)$ is the subsystem fluctuations. $\hat{\mathcal{S}^z_i}$ is defined in Eq.~(\ref{eq:sgn_Sz}), whose definition maintains uniform orientation for all the sites in the initial state and  thereby prevents the cancellation of odd-order cumulants in dynamical evolution.

Using the DW initial state as an example, we compute the first- [Fig.~\ref{fig:Sfig_subsystem_scaling}(a)], second- [Fig.~\ref{fig:Sfig_subsystem_scaling}(d)], and third-order cumulants [Fig.~\ref{fig:Sfig_subsystem_scaling}(g)] using a system size of $L=96$ and subsystem dimensions ranging from $N=10$ to $N=90$. Note here that we select symmetric subsystems centered at the lattice midpoint to suppress the  disturbances from boundary reflections~\cite{wienand2024emergence}.
While spin fluctuations are initially zero for a DW initial state, these nonlocal quantities develop over time as a result of the subsystem being entangled with its environment. At long time, these fluctuations reach a universal regime characterized by a certain scaling exponent.
To extract the corresponding exponents $\nu_n$, we normalize both the local observation values and fluctuations by subsystem size $N$, and identify the threshold time $\tau$ when the local observation values and fluctuations reach a characteristic value of 0.15 [Fig.~\ref{fig:Sfig_subsystem_scaling}(b)(e)(h)]. The scaling law $N \propto \tau^{\nu_n}$ then allows us to determine $\nu_n$ [Fig.~\ref{fig:Sfig_subsystem_scaling}(c)(f)(i)]. 
Our results confirm that this method yields identical scaling exponents of the first few cumulants with those from spin polarization transfer (Figs.~2 and 3 in the main text).

\end{document}